\documentclass[11pt,a4paper]{article}
\pdfoutput=1
\usepackage{jheppub}
\usepackage{amsmath,epsfig}
\usepackage{mathrsfs}
\usepackage{amssymb}
\usepackage{mathtools}
\usepackage{graphics}
\usepackage{cancel}
\usepackage{slashed}
\usepackage{url}
\usepackage{hyperref}

\usepackage{color}
\usepackage[normalem]{ulem}

\newcommand{\be}{\begin{equation}}
\newcommand{\ee}{\end{equation}}
\newcommand{\bea}{\begin{eqnarray}}
\newcommand{\eea}{\end{eqnarray}}

\title{Testable Baryogenesis in Seesaw Models }

\preprint{IFIC/16-38\\ }

\author[a]{ P. ~Hern\'andez,}
\author[a]{M.~Kekic,}
\author[b]{J. ~L\'opez-Pav\'on,}
\author[a]{ J.~ Racker,}
\author[a]{J.~ Salvado.}

\affiliation[a]{Instituto de F\'{\i}sica Corpuscular, Universidad de Valencia and CSIC, 
 Edificio Institutos Investigaci\'on, Catedr\'atico Jos\'e Beltr\'an 2, 46980 Spain}
\affiliation[b]{INFN, Sezione di Genova, via Dodecaneso 33, 16146 Genova, Italy}
\abstract{We revisit the production of baryon asymmetries in the minimal type I seesaw model with heavy Majorana singlets  in the GeV range. In particular we include ``washout'' effects from scattering processes with gauge bosons, Higgs decays and inverse decays, besides the dominant top scatterings. We show that in the minimal model with two singlets, and for an inverted light neutrino ordering, future measurements from SHiP and neutrinoless double beta decay could in principle provide sufficient information  to predict the matter-antimatter asymmetry in the universe. We also show that SHiP measurements could provide very valuable information on the PMNS CP phases.}


\begin{document}

\maketitle
\section{Introduction}

It is well known that minimal extensions of the Standard Model that accommodate massive neutrinos, such as the type I seesaw models, could also explain the observed matter-antimatter asymmetry in the universe \cite{Fukugita:1986hr}. The two new ingredients that make this possible are the existence of new particles that are not in thermal equilibrium sometime before the electroweak phase transition ($T_{EW}\simeq140$GeV) and the presence of new CP-violating interactions of these particles. 

Two basic scenarios have been shown to work. In the first one, the heavy Majorana
singlets decay out of equilibrium generating a lepton asymmetry that sphaleron processes recycle into a baryonic
one. These neutrinos have masses well above the electroweak scale, typically $M \gtrsim 10^8-10^9$~GeV~\cite{Davidson:2002qv,Hambye:2003rt} and  $M \gtrsim 10^6$~GeV when $B-L$ is almost conserved~\cite{Racker:2012vw}, while in resonant leptogenesis masses in the TeV scale are possible~\cite{Pilaftsis:2003gt}. 
For a comprehensive review 
and references of this very well studied scenario see ref.~\cite{Davidson:2008bu}. In the second scenario, the heavy Majorana singlets have masses below the electroweak scale, and therefore their Yukawa couplings are small enough that  one or more of these  states might not reach 
thermal equilibrium by the time the electroweak phase transition takes place. A lepton (and baryon) asymmetry can be generated 
 when the states are being produced, i.e. at freeze-in. The sterile states get populated via Yukawa interactions, but the coherence between collisions 
is essential to produce a CP asymmetry, via the interference of CP-odd and CP-even phases. This is why this mechanism is often 
referred-to as baryogenesis from neutrino oscillations. It 
was first proposed by Akhmedov, Rubakov and Smirnov  (ARS) in their pioneering work \cite{Akhmedov:1998qx} and pursued, with important refinements in references~ \cite{Asaka:2005pn,Shaposhnikov:2008pf}. A list of recent references is  \cite{Asaka:2011wq,Canetti:2012kh,Drewes:2012ma,Shuve:2014zua,Abada:2015rta,Hernandez:2015wna}.

In \cite{Hernandez:2015wna}, we studied this second scenario and explored the available phase space for successful leptogenesis in the minimal models with two or three extra singlets, $N=2, 3$. In particular we considered an accurate analytical approximation, where we could identify the relevant CP invariants, and that helped us explore the full parameter space.  The case with $N=2$ is effectively equivalent to the popular $\nu$MSM\cite{Asaka:2005pn}, which is a $N=2+1$ model, where the lighter neutrino plays the role of dark matter and decouples from the problem in the generation of the baryon asymmetry. The original ARS scenario on the other hand required the interplay of all the three species. The analytical approximation of \cite{Hernandez:2015wna} allowed to clarify these different scenarios. 

The purpose of this paper is twofold. First, we will refine our previous  study in various aspects. In \cite{Hernandez:2015wna} (like in most previous works) the collision terms only included the dominant top quark scatterings.  As has been known for sometime \cite{Anisimov:2010gy,Besak:2012qm}, scatterings off gauge bosons,  as well as the 
 resumed decays and inverse decay processes, are  also very important. These rates have been computed in \cite{Besak:2012qm,Ghisoiu:2014ena} in the limit of 
 vanishing leptonic chemical potentials.  In the generation of lepton asymmetries, it is very important however to include the effect of the latter, since these will tend to washout the asymmetry. In  section 2 we derive new kinetic equations including all the scattering processes considered in \cite{Besak:2012qm,Ghisoiu:2014ena}, that we have re-evaluated in the presence of small leptonic chemical potentials. Furthermore Fermi-Bose statistics is consistently used through-out. 

The second important improvement  concerns our scans of parameter space. In our previous study we speeded-up the scan using the analytical approximation. This forced us to 
avoid some regions in parameter space, to ensure that the approximation was good enough. We have now  optimized significantly the numerical solution of the kinetic equations, in particular addressing the stiffness problem. The  analytical approximation is no longer needed, and therefore the ad hoc constraints on parameter space are avoided. We use a Bayesian approach to extract posterior probabilities on the relevant observables of the model, from a prediction of the measured baryon asymmetry, using the
Multinest package. In section 3, we present the results of these scans of parameter space. 

In the second part of the paper, we address the question: to what extent it would be possible to predict quantitatively the baryon asymmetry, within the minimal model $N=2$,  if the heavy neutrino states would be discovered in future experiments, such as SHiP.  In section 4, we derive 
approximate analytical formulae  valid in the range within SHiP reach, which demonstrate the complementarity of the different measurements: mixings and masses of the extra states from direct searches, neutrinoless double beta decay and the CP phase measurable in neutrino oscillations. 
The numerical study confirms these expectations and allows us to answer the question in the affirmative if nature
is kind enough to provide us with positive signals at SHiP and an inverted neutrino ordering. Furthermore we show how such SHiP measurements could constrain the 
CP-violating phases of the PMNS matrix.

\section{Kinetic equations}

The Lagrangian of the model is given by:
   \begin{eqnarray}
{\cal L} = {\cal L}_{SM}- \sum_{\alpha,i} \bar L^\alpha Y^{\alpha i} \tilde\Phi N^i_R - \sum_{i,j=1}^3 {1\over 2} \bar{N}^{ic}_R M^{ij} N_R^j+ h.c., \nonumber
\label{eq:lag}
\end{eqnarray}
where $Y$ is a $3\times 3$ complex matrix and $M$ a symmetric matrix. One convenient parametrization  is in terms of the eigenvalues of the $Y$  and  $M$ matrices, together with two unitary matrices, $V$ and $W$. In the basis where the Majorana mass is diagonal, $M = {\rm Diag}(M_1,M_2,M_3)$, the neutrino Yukawa matrix is given by:
\begin{eqnarray}
Y \equiv V^\dagger {\rm Diag}(y_1,y_2,y_3) W. 
\label{eq:yuk}
\end{eqnarray} 
Without loss of generality, using rephasing invariance, we can reduce the unitary matrices to the  form:
\begin{eqnarray}
W &=& U(\phi_{12},\phi_{13}, \phi_{23},d)^\dagger {\rm Diag}(1, e^{i \alpha_1}, e^{i \alpha_2}),\nonumber\\
V &=& {\rm Diag}(1, e^{i \beta_1}, e^{i \beta_2}) U(\bar{\phi}_{12},\bar{\phi}_{13}, \bar{\phi}_{23},\bar{d}), 
\end{eqnarray}
where\footnote{Note the unconventional ordering of the 2$\times$2 rotation matrices in $U$.}
\begin{footnotesize}
\begin{eqnarray}
U(\alpha,\beta,\gamma,\delta) \equiv \left( \begin{array}{ccc}
\cos\alpha & \sin\alpha & 0 \\
-\sin\alpha & \cos\alpha & 0 \\
0& 0 &1\end{array}\right)
\left( \begin{array}{ccc}
\cos\beta & 0& \sin\beta e^{-i\delta}\\ 
0 & 1 & 0 \\
-\sin\beta e^{i� \delta} & 0 &\cos\beta\\
 \end{array}\right) 
\left( \begin{array}{ccc}
1 & 0 & 0 \\
0& \cos\gamma & \sin\gamma \\
0& -\sin\gamma  &\cos\gamma\\
\end{array}\right).
\end{eqnarray}
\end{footnotesize}
Obviously not all the parameters are free, since this model must reproduce the light neutrino masses, which approximately implies the seesaw relation:
\begin{eqnarray}
m_\nu \simeq -{v^2\over 2} Y {1\over M} Y^T ,
\end{eqnarray}
 where $v=246$~GeV is the vev of the Higgs. A very convenient parametrization that takes this constraint into account is the Casas-Ibarra one \cite{Casas:2001sr}, where the Yukawa matrix can be written in terms of the light neutrino masses and mixings as
\begin{eqnarray}
Y= -i U^*_{\rm PMNS} \sqrt{m_{\rm light}} R(z_{ij})^T \sqrt{M} {\sqrt{2}\over v},
\label{eq:yci}
\end{eqnarray}
where $m_{\rm light}$ is a diagonal matrix of the light neutrino masses, $U_{\rm PMNS}(\theta_{12}, \theta_{13},\theta_{23},\delta,\phi_1,\phi_2)$ is the PMNS matrix that describes the light neutrino mixing, $M$ is the diagonal matrix of the heavy neutrino masses, and $R$  is a   complex orthogonal matrix,  that depends generically on one (three) complex angle(s) $z_{ij}$ for $N=2$ ($N=3$). 

The kinetic equations that describe the production of sterile neutrinos in the early Universe have been studied in many previous works, see for example \cite{Canetti:2012kh,Drewes:2012ma,Shuve:2014zua}.  
In this work we have rederived these equations with the following refinements with respect to our previous work \cite{Hernandez:2015wna}:
\begin{itemize}
\item Fermi-Dirac or Bose-Einstein statistics is kept throughout
\item Collision terms include $2\leftrightarrow 2$ scatterings at tree level with top quarks and gauge bosons, as well as $1 \leftrightarrow 2$ scatterings including the resummation  of scatterings mediated by soft gauge bosons as obtained in refs. \cite{Anisimov:2010gy,Besak:2012qm,Ghisoiu:2014ena}
\item Leptonic chemical potentials are kept in all collision terms to linear order
\item Include spectator processes
\end{itemize}
As usual we assume that all the spectator particles are in kinetic equilibrium. 
On the other hand, we neglect the effects of the top quark and Higgs chemical potentials. These effects are expected to be smaller than 
the effect of thermal masses in $2 \leftrightarrow 2$ processes that we are neglecting. Note that, in contrast with the  effects of the lepton chemical potential, the former do not bring in any new flavour structure. 
 
The starting point to derive the equations is the Raffelt-Sigl formalism \cite{Sigl:1992fn}, where the sterile neutrino density satisfies the equation: 
\begin{eqnarray}
{d \rho_N(k) \over d t} = -i [H, \rho_N(k)] -{1 \over 2} \left\{ \Gamma^a_N, \rho_N \right\} + {1 \over 2} \left\{ \Gamma^p_N, 1-\rho_N \right\}, 
\end{eqnarray}
where 
\begin{eqnarray}
H \equiv {M^2 \over 2 k_0} + V_N(k), \;\;\; V_N(k) \equiv {T^2 \over 8 k_0} Y^\dagger Y ,
\end{eqnarray}
and $\Gamma^a_N(k)$ and $\Gamma^p_N(k)$ are the annihilation and production rates of the sterile neutrinos.

The result can be written  as
\begin{eqnarray}
\Gamma^{p}_{N ij} &=& Y^\dagger_{i\alpha} \rho_F\left({k_0\over T}- \mu_\alpha\right) \gamma_{N }(k,\mu_\alpha) Y_{\alpha j},\nonumber\\
\Gamma^{a}_{N ij} &=& Y^\dagger_{i\alpha}
\left(1-\rho_F\left({k_0\over T}- \mu_\alpha\right)\right) \gamma_{N
}(k,\mu_\alpha) Y_{\alpha j} ,
\end{eqnarray}
where $\rho_F(y) =( \exp y + 1)^{-1}$ is the Fermi-Dirac distribution and $\mu_\alpha$ is the leptonic chemical potential normalised by the temperature.  $\gamma_N$ contain the contributions from all  $2 \rightarrow 2$ processes that produce an $N$:
\begin{eqnarray}
\bar{Q} t \rightarrow \bar{l} N;\; t l \rightarrow Q N;\; \bar{Q} l
\rightarrow \bar{t} N;\; W l \rightarrow \bar{\phi} N;\; l \phi \rightarrow
W N;\; W \phi \rightarrow \bar{l} N ,
\end{eqnarray}  
and $1\leftrightarrow 2$ processes: $\phi \rightarrow \bar{l} N$ including resummed soft-gauge interactions. 
All these contributions have been computed for vanishing leptonic chemical potential in \cite{Asaka:2011wq,Besak:2012qm,Ghisoiu:2014ena}. We have followed their methods including the effects of a  lepton chemical potential  to linear order. 

\begin{figure}
 \begin{center}
\includegraphics[scale=0.42]{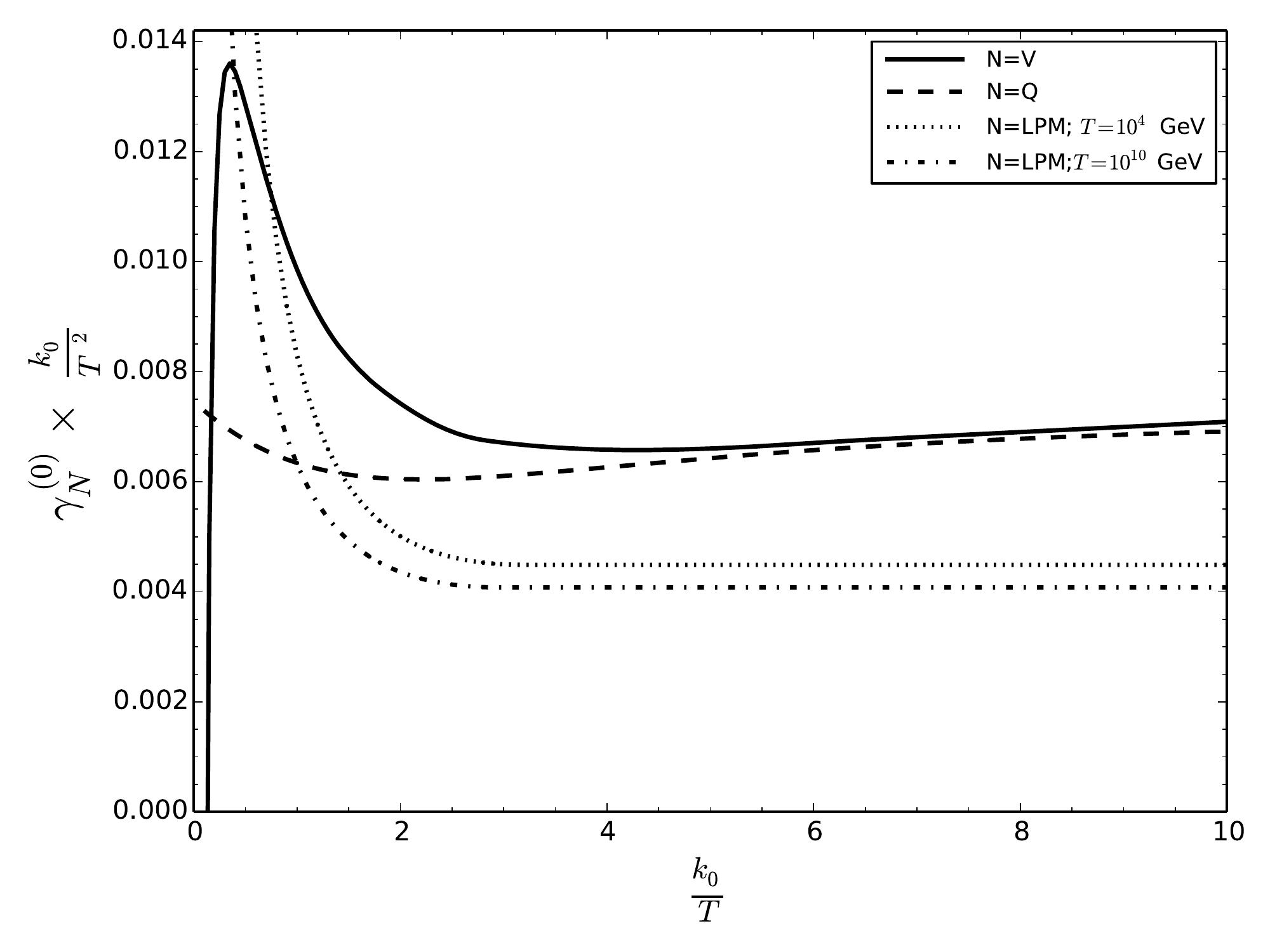} \includegraphics[scale=0.42]{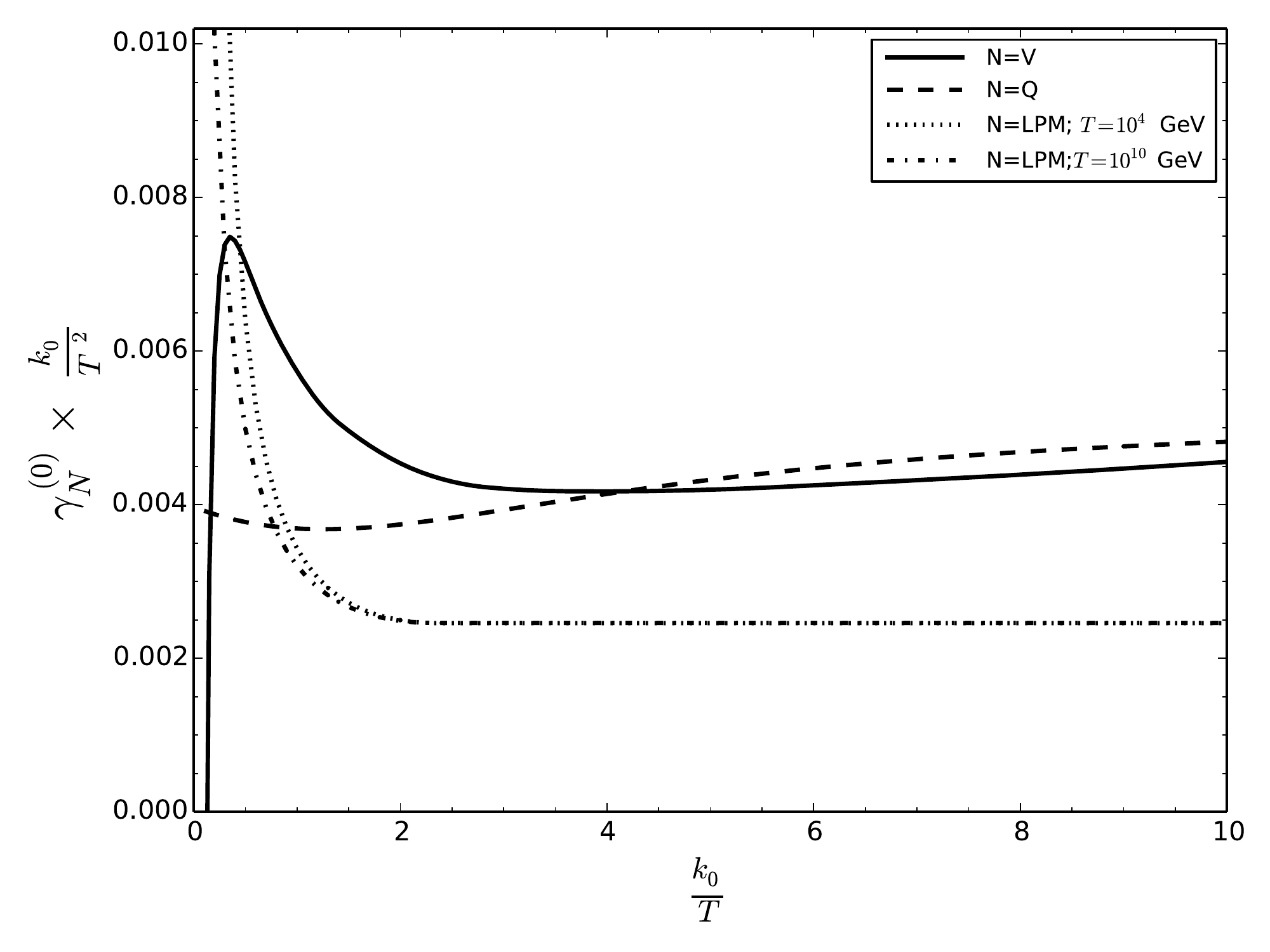} \includegraphics[scale=0.42]{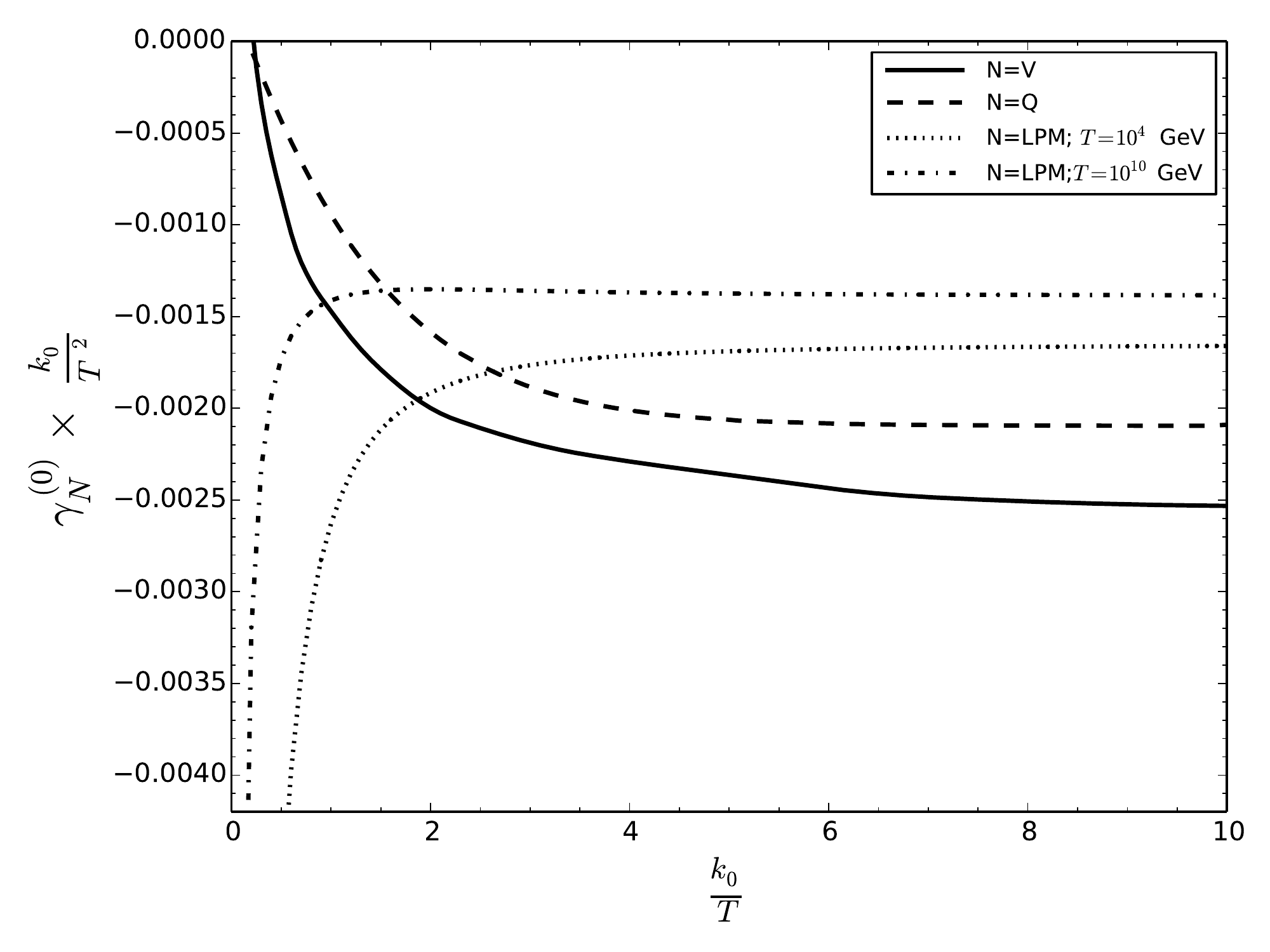} 
\caption{\label{fig:gammas} $\gamma_V^{(i)} k_0/T^2$ (solid) $\gamma_Q^{(i)} k_0/T^2$ (dashed), $\gamma_{LPM}^{(i)}(k_0) k_0/T^2$ for $T=10^4$ GeV (dotted) 
and $\gamma_{LPM}^{(i)}(k_0) k_0/T^2$ for $T=10^{10}$ GeV (dash-dotted) as a function of $k_0/T$. }
\end{center}
\end{figure}

Defining 
\begin{eqnarray}
\gamma_N(k, \mu_\alpha) \simeq \gamma^{(0)}_N(k) + \gamma^{(2)}_N(k) \mu_\alpha, 
\end{eqnarray}
and
\begin{eqnarray}
\gamma_N^{(1)} \equiv  \gamma_N^{(2)} -{\rho'_F \over \rho_F} \gamma_N^{(0)},
\end{eqnarray}
with $\rho'_F(y) \equiv {d \rho_F(y)\over dy}$, the functions $\gamma_N^{(i)}$ get contributions from quark (Q), gauge scattering (V) and the $1\rightarrow 2$ resummed processes (LPM):
\begin{eqnarray}
\gamma_N^{(i)} = \gamma_{LPM}^{(i)}+ y_t^2 \gamma_{Q}^{(i)} +  (3 g^2+ g'^2) \left( \gamma_{V}^{(i)} + \gamma_{IR}^{(i)} \log\left({1 \over 3 g^2+ g'^2}\right)\right).
\end{eqnarray}
The functions $\gamma_{Q,V}^{(i)}$ depend only on the ratio $k_0/T$, while $\gamma_{LPM}^{(i)}$ has non-trivial temperature dependence due to the runnings of the 
coupling constants\footnote{For the details of the calculation see \cite{Ghisoiu:2014ena}.}. In Fig.~\ref{fig:gammas} the three functions are plotted, where the two lines labeled LPM curve correspond to two temperatures   
$10^4$ GeV and $10^{10}$ GeV, while 
\begin{eqnarray}
\gamma_{IR}^{(0)} = 2 \gamma_{IR}^{(1)} = {T^2\over 256 \pi k_0} {\rho_B\over \rho_F}, \;\;\;\gamma_{IR}^{(2)} = \gamma_{IR}^{(0)} {\rho'_F\over \rho_F} + \gamma_{IR}^{(1)},
\end{eqnarray}
where $\rho_B(y)= (\exp y - 1)^{-1}$ is the Bose-Einstein distribution.

Inserting these functions in the kinetic equation we get:
\begin{eqnarray}
{d \rho_N \over d t} &=& -i [H, \rho_N] -{ \gamma_N^{(0)}\over 2} \left\{ Y^\dagger Y  , \rho_N-\rho_F \right\} + \gamma_N^{(1)}  \rho_F  Y^\dagger \mu Y  \nonumber\\
& &- { \gamma_N^{(2)}\over 2} \left\{ Y^\dagger \mu Y  , \rho_N\right\}, 
\end{eqnarray}
where $\mu \equiv {\rm Diag}(\mu_\alpha)$. 
The equation for the antineutrino (opposite helicity state) density is the same but changing $\mu \rightarrow -\mu$ and $Y\rightarrow Y^*$. 

It is often useful  to consider instead the evolution of the CP conserving and violating combinations:
\begin{eqnarray}
\rho_\pm \equiv {\rho_N \pm\rho_{\bar N} \over 2}. 
\end{eqnarray}
The equations for these combinations are:
\begin{eqnarray}
\dot{\rho}_+ &=& -i [H_{\rm re}, \rho_+] +  [ H_{\rm im}, \rho_-] -{\gamma^{(0)}_N\over 2} \big\{ {\rm Re}[Y^\dagger Y], \rho_+-\rho_{F}\big\} \nonumber\\
&&+ i \gamma_N^{(1)} {\rm Im}[Y^\dagger \mu Y]  \rho_{F}  - i {\gamma_N^{(2)} \over 2}  \big\{{\rm Im}[Y^\dagger \mu Y],\rho_+\big\}-i {\gamma_N^{(0)}\over 2} \left\{{\rm Im}[Y^\dagger Y], \rho_-\right\},\nonumber\\  
\dot{\rho}_- &=& -i [H_{\rm re}, \rho_-] +  [ H_{\rm im}, \rho_+] -{\gamma_N^{(0)}\over 2} \big\{{\rm Re}[Y^\dagger Y], \rho_-\big\}\nonumber\\
&&+  \gamma_N^{(1)} {\rm Re}[Y^\dagger \mu Y]  \rho_{F}  -  {\gamma^{(2)}_N \over 2}  \big\{{\rm Re}[Y^\dagger \mu Y],\rho_+\big\}-i {\gamma_N^{(0)}\over 2} \left\{{\rm Im}[Y^\dagger Y], \rho_+-\rho_{F}\right\}.\nonumber\\
\label{eq:rhoprhom}
\end{eqnarray}

Finally we need the equations that describe the evolution of the leptonic chemical potentials. This is obtained from the equation 
that describes the evolution of the conserved charges in the absence of neutrino Yukawas, that is the ${B \over 3} -L_\alpha$ numbers. 
These numbers can only be changed by the same out of equilibrium processes that produce the sterile neutrinos: 
\begin{eqnarray}
{d n_{B/3 -L_\alpha} \over d t} = {1 \over 2} \int_p \left\{ \Gamma^a_l(p), \rho_l(p,\mu) \right\}_{\alpha\alpha} - {1 \over 2} \int_p \left\{ \Gamma^p_l(p), 1-\rho_l(p,\mu) \right\}_{\alpha\alpha}. 
\end{eqnarray}
where $\int_p \equiv \int {d^3 p\over (2 \pi)^3}$. 
Since $(\rho_l)_{\alpha\alpha}= \rho_F\left({p_0\over T}-\mu_\alpha\right)$, 
it is possible to relate the integrated rates of these equations to those of the sterile neutrinos and their densities:
\begin{eqnarray}
\dot{n}_{B/3-L_\alpha} & = & 
-2 \int_{k}  \left\{{\gamma_N^{(0)}\over 2} (Y \rho_N Y^\dagger- Y^* \rho_{\bar N} Y^T)_{\alpha\alpha} \right.\nonumber\\
&+&\left.\mu_\alpha \left({\gamma_N^{(2)}\over 2} (Y \rho_N Y^\dagger+Y^* \rho_{\bar N} Y^T)_{\alpha\alpha} - \gamma_N^{(1)} {\rm Tr}[YY^\dagger I_\alpha] \rho_{F}  \right)\right\},
\label{eq:bml}
\end{eqnarray}
or in terms of $\rho_\pm$:
\begin{eqnarray}
\dot{n}_{B/3-L_\alpha} & = & 
-2 \int_{k}  \left\{ \gamma_N^{(0)} {\rm Tr}[\rho_- {\rm Re}(Y^\dagger  I_\alpha Y) + i\rho_+  {\rm Im}(Y^\dagger I_\alpha Y)]\right.\nonumber\\
&+&\left.\mu_\alpha \left(\gamma_N^{(2)} {\rm Tr}[\rho_+ {\rm Re}(Y^\dagger I_\alpha Y)] - \gamma_N^{(1)} {\rm Tr}[YY^\dagger I_\alpha] \rho_{F}  \right)\right\},
\label{eq:bml2}
\end{eqnarray}
where $I_\alpha$ is the projector on flavour $\alpha$, and we have neglected terms of ${\mathcal O}(\mu \rho_-)$. 

The relation between the leptonic chemical potentials and the approximately conserved charges, $B/3 - L_\alpha$, is given for $T \leq 10^6$ GeV by \cite{Nardi:2006fx}
\begin{eqnarray}
\mu_\alpha = - \sum_\beta C_{\alpha\beta} \mu_{B/3-L_\beta}, \;\; C_{\alpha\beta} = {1\over 711} \left(\begin{array}{lll} 221 & -16 & -16\\
-16 & 221 & -16 \\
-16 & -16 & 221\end{array}\right),
\label{eq:chem}
\end{eqnarray}
where we have defined $\mu_{B/3-L_\beta}$ by the relation:
\begin{eqnarray}
n_{B/3-L_\alpha} \equiv -2 \mu_{B/3 -L_\alpha} \int_k \rho'_F = {1\over 6} \mu_{B/3 -L_\alpha} T^3
.
\end{eqnarray}

Introducing finally the expansion of the universe and changing variables to the scale factor $x=a$ and $y=k a$, the time derivative of the distribution functions change to:
\begin{eqnarray}
{d\rho_N(T,k) \over d t} \rightarrow \left. x H_u(x) {\partial \rho_N(x,y)\over\partial x}\right|_{\rm y ~fixed}\nonumber\\
{d n_{B/3 -L_\alpha}\over d t} \rightarrow -2 x H_u(x) {d\mu_{B/3 - L_\alpha}\over d x} \int_k \rho'_F,
\end{eqnarray}
where $H_u(x)= \sqrt{4 \pi^3 g_*(T)\over 45} {T^2\over M_{\rm Planck}}$ is the Hubble expansion parameter. Assuming a radiation dominated universe with constant number of relativistic degrees of freedom $g_*(T_0)\simeq 106.75$ for $T_0 \geq T_{EW}$, then $x T$= constant that we can fix to one. 

\subsection{Momentum averaging}

In principle the equations should be solved for all momenta of the sterile neutrinos, but it is a good approximation \cite{Asaka:2011wq} to assume 
$\rho_\pm(x,y) = r_\pm(x) \rho_F(y)$, with $r_\pm(x)$ independent of momentum. This allows to integrate explicitly over $y$ and we get
the momentum-averaged equations:
\begin{eqnarray}
x H_u {d r_+\over d x} &=& -i [\langle H_{\rm re}\rangle, r_+] +  [ \langle H_{\rm im}\rangle, r_-] -{\langle\gamma^{(0)}_N\rangle\over 2} \{{\rm Re}[Y^\dagger Y], r_+-1\} \nonumber\\
&&+ i \langle \gamma_N^{(1)} \rangle {\rm Im}[Y^\dagger \mu Y]   - i {\langle \gamma_N^{(2)}\rangle \over 2}  \big\{{\rm Im}[Y^\dagger \mu Y],r_+\big\}-i {\langle\gamma_N^{(0)}\rangle\over 2} \left\{{\rm Im}[Y^\dagger Y], r_-\right\},\nonumber\\  
x H_u {d r_-\over d x} &=& -i [ \langle H_{\rm re}\rangle, r_-] +  [\langle H_{\rm im}\rangle, r_+] -{\langle \gamma_N^{(0)}\rangle\over 2} \big\{{\rm Re}[Y^\dagger Y], r_-\big\}\nonumber\\
&&+  \langle\gamma_N^{(1)}\rangle {\rm Re}[Y^\dagger \mu Y]   -  {\langle \gamma^{(2)}_N \rangle \over 2}  \big\{{\rm Re}[Y^\dagger \mu Y], r_+\big\}-i {\langle\gamma_N^{(0)}\rangle\over 2} \left\{{\rm Im}[Y^\dagger Y], r_+-1\right\},\nonumber\\
x H_u {d{\mu}_{B/3-L_\alpha}\over d x} & = & {\int_k \rho_F\over \int_k \rho'_F}  \Big\{\langle\gamma_N^{(0)}\rangle {\rm Tr}[r_- {\rm Re}(Y^\dagger  I_\alpha Y) + i r_+  {\rm Im}(Y^\dagger I_\alpha Y)]\nonumber\\
&+ &\mu_\alpha \left(\langle\gamma_N^{(2)}\rangle {\rm Tr}[r_+ {\rm Re}(Y^\dagger I_\alpha Y)] - \langle \gamma_N^{(1)}\rangle {\rm Tr}[YY^\dagger I_\alpha]  \right)\Big\},\nonumber\\
\mu_{\alpha} &=& - \sum_\beta C_{\alpha\beta} \mu_{B/3-L_\beta},
\label{eq:rhoprhomav}
\end{eqnarray}
where
\begin{eqnarray}
\langle (...)\rangle \equiv {\int_y (...) n_F(y)\over \int_y n_F(y)}
\end{eqnarray}
and ${\int_k \rho_F\over \int_k \rho'_F} = -{9 \xi(3)\over \pi^2}$.

The equation for $r_N = \rho_N/\rho_F$ and $r_{\bar N} = \rho_{\bar N}/\rho_F$ are equivalently
\begin{eqnarray}
x H_u {d r_N\over d x} &=& -i [\langle H\rangle, r_N]  -{\langle\gamma^{(0)}_N\rangle\over 2} \{Y^\dagger Y, r_N-1\}
+  \langle \gamma_N^{(1)} \rangle Y^\dagger \mu Y   -  {\langle \gamma_N^{(2)}\rangle \over 2}  \big\{Y^\dagger \mu Y,r_N\big\},\nonumber\\ 
x H_u {d r_{\bar N}\over d x} &=& -i [\langle H^*\rangle, r_{\bar N}]  -{\langle\gamma^{(0)}_N\rangle\over 2} \{Y^T Y^*, r_{\bar N}-1\} 
-  \langle \gamma_N^{(1)} \rangle Y^T \mu Y^*   +  {\langle \gamma_N^{(2)}\rangle \over 2}  \big\{Y^T \mu Y^*,r_{\bar N}\big\},\nonumber\\  
x H_u {d{\mu}_{B/3-L_\alpha}\over d x} & = & {\int_k \rho_F\over \int_k \rho'_F}  \left\{{\langle \gamma_N^{(0)\rangle}\over 2} (Y r_N Y^\dagger- Y^* r_{\bar N} Y^T)_{\alpha\alpha} \right.\nonumber\\
&+&\left.\mu_\alpha \left({\langle\gamma_N^{(2)}\rangle\over 2} (Y r_N Y^\dagger+Y^* r_{\bar N} Y^T)_{\alpha\alpha} - \langle\gamma_N^{(1)}\rangle {\rm Tr}[YY^\dagger I_\alpha]   \right)\right\},\nonumber\\
\mu_{\alpha} &=& - \sum_\beta C_{\alpha\beta} \mu_{B/3-L_\beta}.
\label{eq:rhonrhonbarav}
\end{eqnarray}

The momentum averaged rates are:
\begin{eqnarray}
\langle \gamma_N^{(i)}\rangle =  A_i  \left[  c^{(i)}_{LPM} +y_t^2 c^{(i)}_Q + (3 g^2+ g'^2) \left(c^{(i)}_V+\log\left({1 \over 3 g^2+ g'^2}\right) \right)\right],
\end{eqnarray}
with 
\begin{eqnarray}
A_0 = 2 A_1 = -4 A_2 \equiv { 4 \pi^2 \over 3 \xi(3)} {T\over 3072 \pi},
\end{eqnarray}
and the coefficients are given in the table~\ref{tab:table}.
\begin{table}
\begin{center}
\begin{tabular}{lllll}
$i$ & $c^{(i)}_{LPM}(T_1)$ & $c^{(i)}_{LPM}(T_2)$ &$c^{(i)}_Q$ & $c^{(i)}_V$\\
\hline
0 & 4.22 &  2.65 & 2.52 & 3.17\\
1 & 3.56 & 2.80 & 3.10 & 3.83\\
2 & 4.77 & 2.50 & 2.27 & 2.89\\
\hline
\end{tabular}
\caption{Coefficients in the momentum averaged rates. The LPM ones have been evaluated at $T_1= 10^4$ GeV and $T_2=10^{10}$ GeV.}
\label{tab:table}
\end{center}
\end{table}

For the couplings $g, g'$ we evaluate them at the scale $\pi T$:
\begin{eqnarray}
{1\over g(\pi T)^2} = {1 \over g(M_z)^2} + {19 \over 48 \pi^2 } \ln\left({\pi T\over M_z}\right),\\
{1 \over g'(\pi T)^2} = {1 \over g'(M_z)^2} + {41 \over 48 \pi^2 } \ln\left({M_z \over \pi T}\right),
\end{eqnarray}
while the top Yukawa running is obtained  numerically from  the one loop renormalization group equations.

\begin{figure}
 \begin{center}
\includegraphics[scale=0.5]{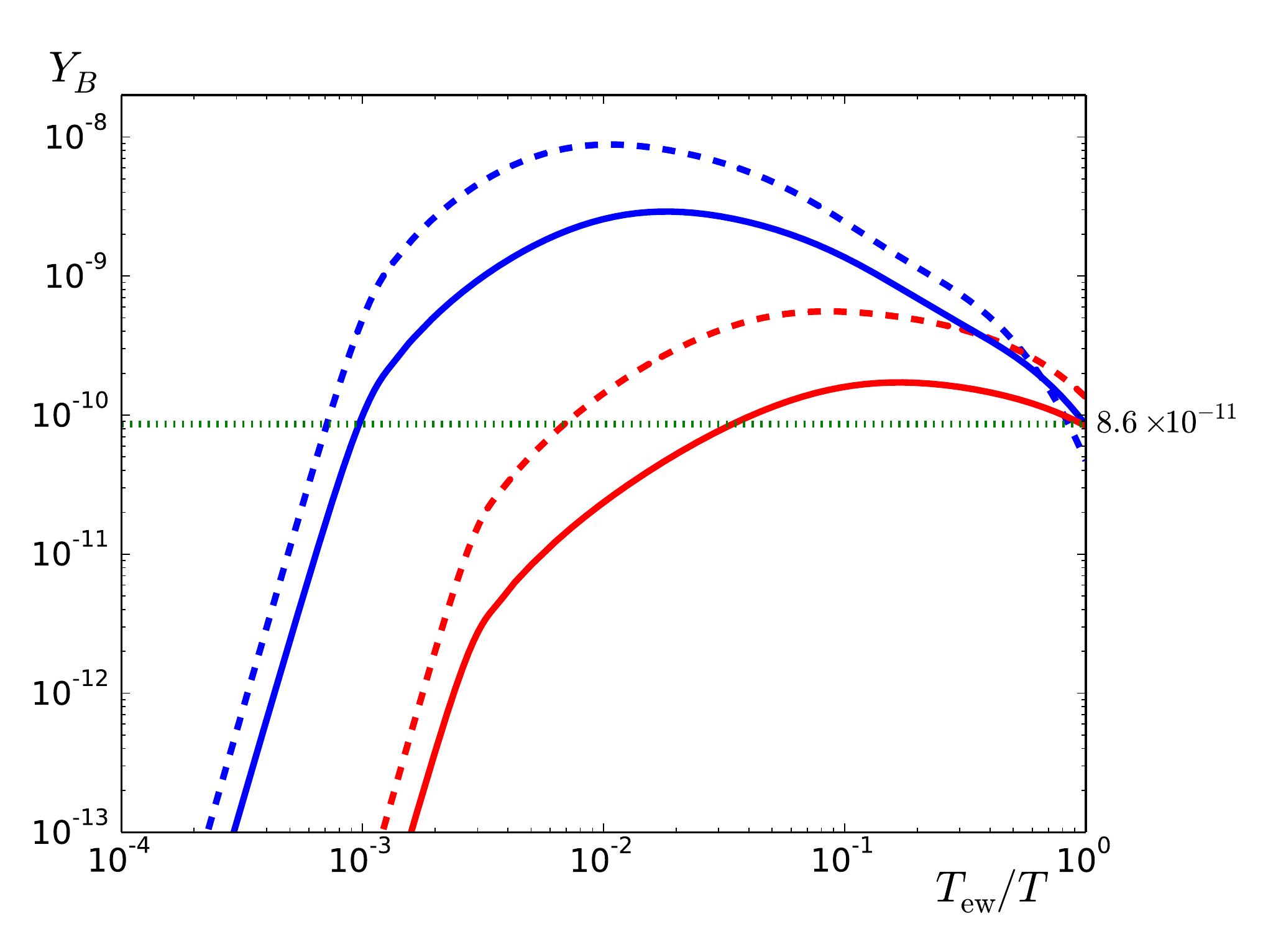} 
\caption{ Time evolution of the baryon asymmetry for two different choices of parameters. First set (blue curves): $z=0.81+3.22 i, \phi_1= 1.21, \delta =2.07, M_1=9.683~$GeV, $M_2 = 9.677$ GeV; second set (red curves): $z=0.88- 0.35 i , \phi_1= 1.65, \delta =-2.07,  M_1=0.754~$GeV, $M_2 = 0.750$ GeV,  using the equations that only take into account quark scattering (solid) and those in eqs.~(\ref{eq:rhonrhonbarav}) (dashed). }
\label{fig:compoldnew} 
\end{center}
\end{figure}

In Fig.~\ref{fig:compoldnew} we compare the time evolution of  the asymmetry that we obtain with the new equations and the old equations of  \cite{Hernandez:2015wna} that only included top scattering processes and Maxwell-Boltzmann statistics. As expected the larger scattering rates induce a larger asymmetry at short times, but also a stronger washout at late times. 

\begin{figure}
 \begin{center}
\includegraphics[scale=0.5]{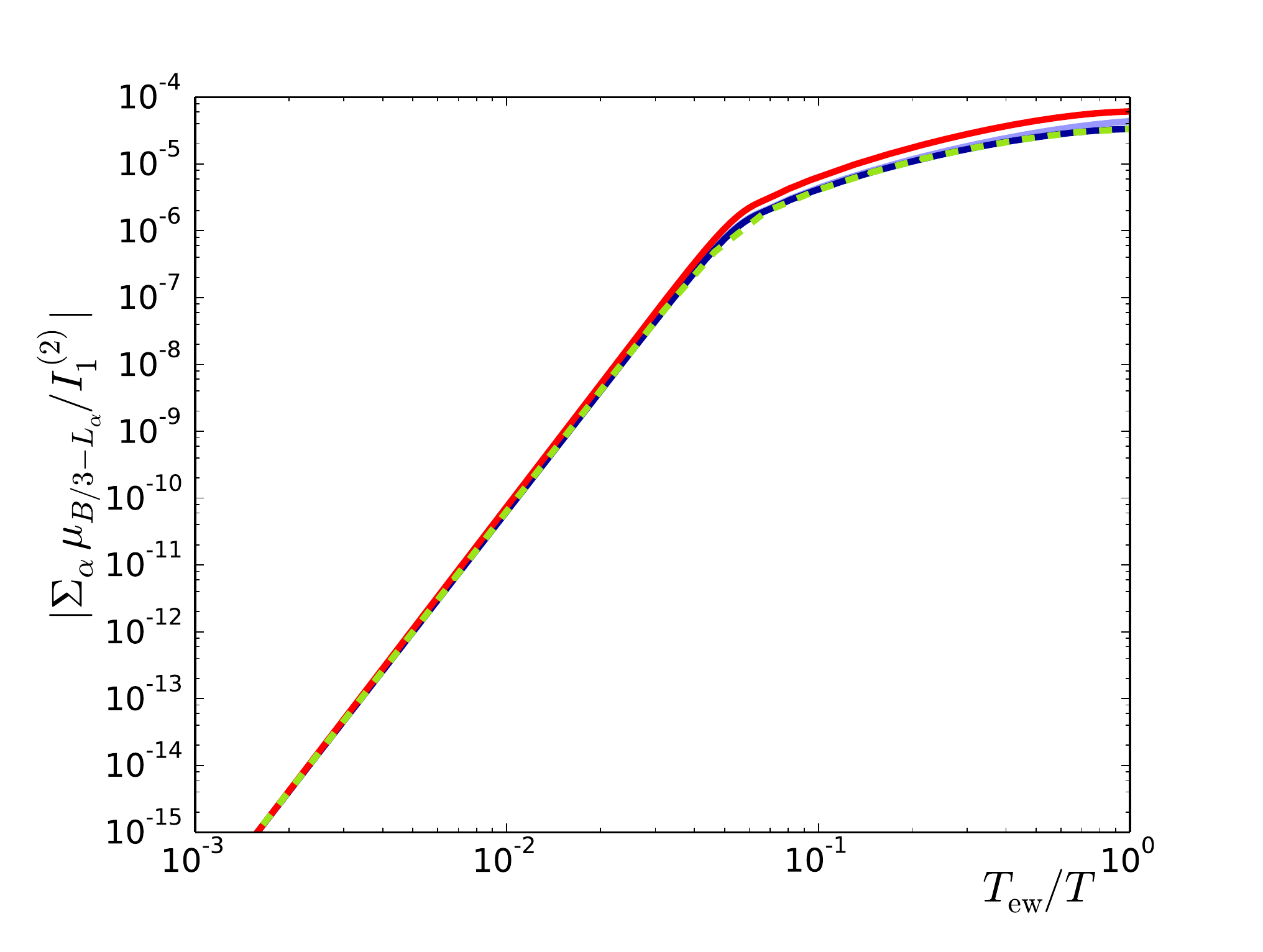} 
\caption{\label{fig:companal}  Assuming parameters for which only the simplest invariant, $I_1^{(2)}$ is non-vanishing, comparison of  the analytical result of eq.(\ref{eq:anal}) (dashed), the full numerical one (red), the numerical one neglecting the variation of the running of the couplings (blue) and that neglecting also non-linear terms (green).  The parameters have been chosen as $M_1= 1$ GeV, $M_2 -M_1 = 10^{-3}$ GeV, and $(y_1,y_2) = (10^{-7}, \sqrt{2} \times 10^{-7})$, and sufficiently small $V, W$ mixings. }
\end{center}
\end{figure}

In \cite{Hernandez:2015wna}, we identified four independent CP rephasing invariants that can contribute to this asymmetry in the general case with $N=3$ as:
\bea 
I_1^{(2)}& =&  -{\rm Im} [W^*_{1 2} V_{1 1} V^*_{2 1}  W_{2 2 } ] ,
\\
I_1^{(3)} &=&   {\rm Im} [W^*_{12} V_{13} V^*_{2 3}  W_{2 2 } ] ,
\\
I_2^{(3)} &=&  {\rm Im} [W^*_{13} V_{12} V^*_{2 2}  W_{2 3 } ],
\nonumber\\
J_W  &=& - {\rm Im}[W^*_{23} W_{22} W^*_{32} W_{33} ].
\label{eq:cpinvs}
\eea
where $V, W$ are the matrices parametrizing the neutrino Yukawa matrix, eq.~(\ref{eq:yuk}). In the minimal scenario with $N=2$ only the first two invariants can contribute.
 We considered a  convenient analytical approximation, based on a perturbative expansion in the mixing angles of these matrices, that allowed us to solve the differential equations analytically, neglecting non-linear terms. It is straightforward to apply the same method to the new equations. As an example we give the result for the asymmetry when  only the CP invariant $I_1^{(2)}$ survives (i.e. for $\phi_{i3} = \bar{\phi}_{i3}  = 0$). Defining
 \begin{eqnarray}
\Delta_{ij} \equiv {\Delta M_{ij}^2\over 2 y} M_P^*, \;\; \Delta_v= (y_2^2-y_1^2) {M_P^* \over 8 y},  \;\; \gamma^{(i)} \equiv \langle \gamma_N^{(i)}\rangle{ M_P^*\over T},
 \end{eqnarray}
 with $M_P^*\equiv M_{\rm Planck} \sqrt{{45\over 4 \pi^3 g_*(T_0)}}$, and neglecting the running of the couplings,  the result is 
  \begin{eqnarray}
  \sum_\alpha \mu_{B/3 -L_\alpha}(t) =  {2 \over 3} \left({9 \xi(3)\over \pi^2}\right)^2 I_1^{(2)}  y_1 y_2 (y_2^2- y_1^2) {(\gamma^{(0)})^2 \gamma^{(1)} \over \bar{\gamma}}   G_1(t),\nonumber\\
  \label{eq:anal}
  \end{eqnarray}
 where
 \begin{eqnarray}
\bar{\gamma}\equiv \sqrt{\left(\gamma^{(0)} + {221\over 711} {9 \xi(3)\over \pi^2} \gamma^{(1)} \right)^2  + {1024 \over  505521} \left({9 \xi(3)\over \pi^2}\right)^2 {y_2^2 y_1^2 \over (y_2^2-y_1^2)^2} (\gamma^{(1)})^2},
\end{eqnarray}
and
   \begin{eqnarray}
G_1(t) &\equiv&   \left(e^{-\Gamma_+  t} - e^{-\Gamma_- t }\right) {\rm Re}\left[i J_{200}(\Delta_{12},-\Delta_{12}, t)+2 \Delta_v J_{201}(\Delta_{12},-\Delta_{12}, t)\right]  \nonumber\\
&+&{1\over 2}  \sum_{\sigma=\pm} \sigma e^{- \Gamma_\sigma t}  {\rm Re}\left[J_{210}(\Delta_{12},-\Delta_{12}, t) \left(-2 \Delta_v + i (2 \Gamma_\sigma - (y_2^2+y_1^2) \gamma_N^{(0)} )\right) \right],   \nonumber\\
\label{eq:g1}
\end{eqnarray}
\begin{eqnarray}
\Gamma_\pm \equiv  {y_1^2+y_2^2\over 2} \left(\gamma^{(0)} + {9 \xi(3)\over \pi^2} {221\over 711} \gamma^{(1)}\right) \mp {y_2^2-y_1^2\over 2} \bar{\gamma}.
\end{eqnarray}
The integrals are
\begin{eqnarray} 
J_{2nm}(\alpha,-\alpha, t) \equiv \int_0^t d x_1~ x_1^n~ e^{i {\alpha x_1^3\over 3}} \int_0^{x_1}~ d x_2 ~x_2^m ~e^{-i {\alpha x_2^3\over 3}}.
\end{eqnarray}
Fig.~\ref{fig:companal}  shows the analytical result compared with the numerical solution to the equation for sufficiently small 
mixing angles.

\subsection{Baryon asymmetry}

The observed baryon asymmetry is usually quoted in terms of the abundance, which is the number-density asymmetry of baryons normalised to the entropy density. After Planck  this quantity is known to per cent precision \cite{Ade:2015xua}:
\begin{eqnarray}
Y_B^{\rm exp} \simeq 8.65(8) \times 10^{-11}.
\end{eqnarray}

  The baryon abundance is related to that of $B-L$  by \cite{Harvey:1990qw,Laine:1999wv}
\begin{eqnarray}
Y_B \simeq {28 \over 79} Y_{B-L}, 
\end{eqnarray}
and 
\begin{eqnarray}
Y_{B-L} = \sum_\alpha {n_{B/3-L_\alpha}\over s}, \;\; n_{B/3-L_\alpha}= {T^3 \over 6} \mu_{B/3-L_\alpha}, \;\; s= {2 \pi^2 \over 45} g_* T^3,
\end{eqnarray}
where   we take $g_* = 106.75$  (which ignores the contribution to the entropy of the sterile states). Our estimate for the baryon asymmetry is therefore
\begin{eqnarray}
Y_B \simeq 1.3 \times 10^{-3}  \sum_\alpha \mu_{B/3 -L_\alpha}.
\label{eq:yb}
\end{eqnarray}

\section{Numerical Results}

The numerical solution of the kinetic equations is challenging, because in most part of the parameter space, they are stiff since
 the oscillation time is much shorter than the collision one. 
In \cite{Hernandez:2015wna} we performed an exploration of the parameter space viable for leptogenesis for the models $N=2,3$ employing the perturbative analytical approximation. This required to constrain certain regions of parameter space where the perturbative solution could fail.  We want to improve on  this scan by going beyond the perturbative estimate of the asymmetry and using the full numerical solution of the equations. 

We have solved eqs.~(\ref{eq:rhonrhonbarav}) using the publicly
available code SQuIDS \cite{Delgado:2014kpa,squids_URL}. The code is designed to solve the evolution
of a generic density matrix in the interaction picture. The interaction picture  is useful because it 
removes the short time scale, i.e. the oscillation scale, from the numerical integration, but fast oscillatory coefficients then
appear in the  terms involving the off-diagonal elements of the density matrix.
 In order to optimize the code,  at some large enough time, we switch to a fully decoherent evolution 
(when the exponents of all the oscillatory terms are larger than $10^5$). 
The decoherent evolution is already included in the
last  version of the SQuIDs code \cite{squids_URL}.
Using this approximation the solution speeds up
the computation by a factor more than a hundred and the result agrees with the full solution with a relative
error smaller than $O(1\%)$. 

Using these optimizations we get a computational time for the full
numerical solution of order minutes, which allows us to
do a Bayesian parameter estimation from the log-likelihood:
\begin{eqnarray}
\log {\mathcal L} = -{1\over 2} \left({Y_B(t_{\rm EW})-Y^{\rm exp}_B\over \sigma_{Y_B}}\right)^2.
\end{eqnarray}
For this, we use a nested sampling algorithm
implemented in the public package MultiNest
\cite{Feroz:2007kg,Feroz:2008xx,Feroz:2013hea} and the Markov Chain sample analysis tool
GetDist \cite{getdist_url} to get the posterior probabilities. The number of random starting points is 5000. 

The scan is performed using the Casas-Ibarra parameters of eq.~(\ref{eq:yci}).  We fix the light neutrino masses and mixings to the present best fit points 
in the global analysis of neutrino oscillation data of ref.~ \cite{Gonzalez-Garcia:2014bfa} for each of the neutrino orderings (normal, NH, and inverted, IH), and leave as free parameters: the complex angle(s) of the $R$ matrix, the CP phases of the PMNS matrix, the lightest neutrino mass as well
as the heavy Majorana masses. For $N=2$ these are six independent parameters, while for $N=3$, there are thirteen free parameters. 

In this work we  consider the  simplest case of $N=2$, which can be obtained from the $N=3$ model in the limit where one of the sterile neutrinos is effectively decoupled, that we can assume without loss of generality to be $N_3$. This can be achieved with the choice of parameters:
\begin{eqnarray}
m_{3(1)}=0, z_{i3} =0, ~R(z_{ij}) \rightarrow R(z_{ij}) (P) 
\label{eq:dec}
\end{eqnarray}
for the IH(NH), where $P$ is the $123\rightarrow 312$ permutation matrix (only necessary for the NH). 

This  is then the model that has been considered in most previous works on the subject \cite{Asaka:2005pn,Shaposhnikov:2008pf,Canetti:2010aw,Asaka:2011wq,Shuve:2014zua}, where the number of constrained parameters is reduced to six: only one complex angle in $R$, $z\equiv \theta+i\gamma$, two CP phases, $\delta$ and $\phi_1$ in the PMNS matrix, and two Majorana neutrino masses, $M_1, M_2$.   

\begin{figure}[h]
 \begin{center}

\includegraphics[scale=0.4]{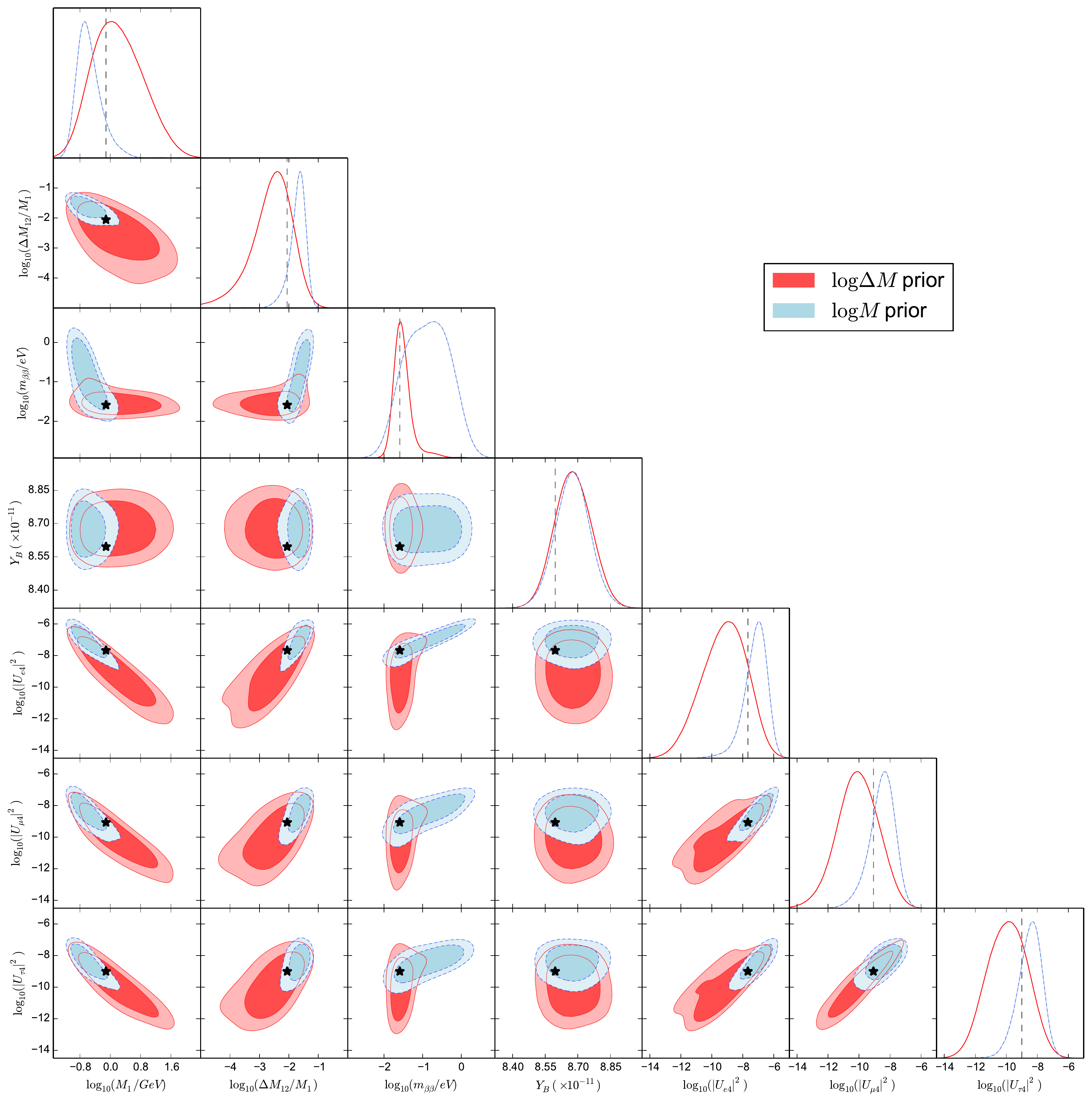} 
\caption{\label{fig:trianglen2ih} Triangle plot with 1D posterior probabilities and  2D $68\%$ and $90\%$ probability contours in the $N=2$ scenario for IH. The parameters shown are the observables $M_1$, $\Delta M_{12}/M_1=(M_2-M_1)/M_1$, $m_{\beta\beta}$, $Y_B$,  and the three mixings with the first of the heavy states $|U_{\alpha 4}|^2$ for $\alpha=e,\mu, \tau$. The blue and red contours correspond respectively to the assumption of a flat prior in $\log_{10} M_1$ and $\log_{10} M_2$ and to a flat prior in $\log_{10} 
M_1$ and $\log_{10}(\Delta M_{12})$. The star is the test point used for the SHiP study of the next section.
 }
\end{center}
\end{figure}

\begin{figure}[h]
 \begin{center}

\includegraphics[scale=0.4]{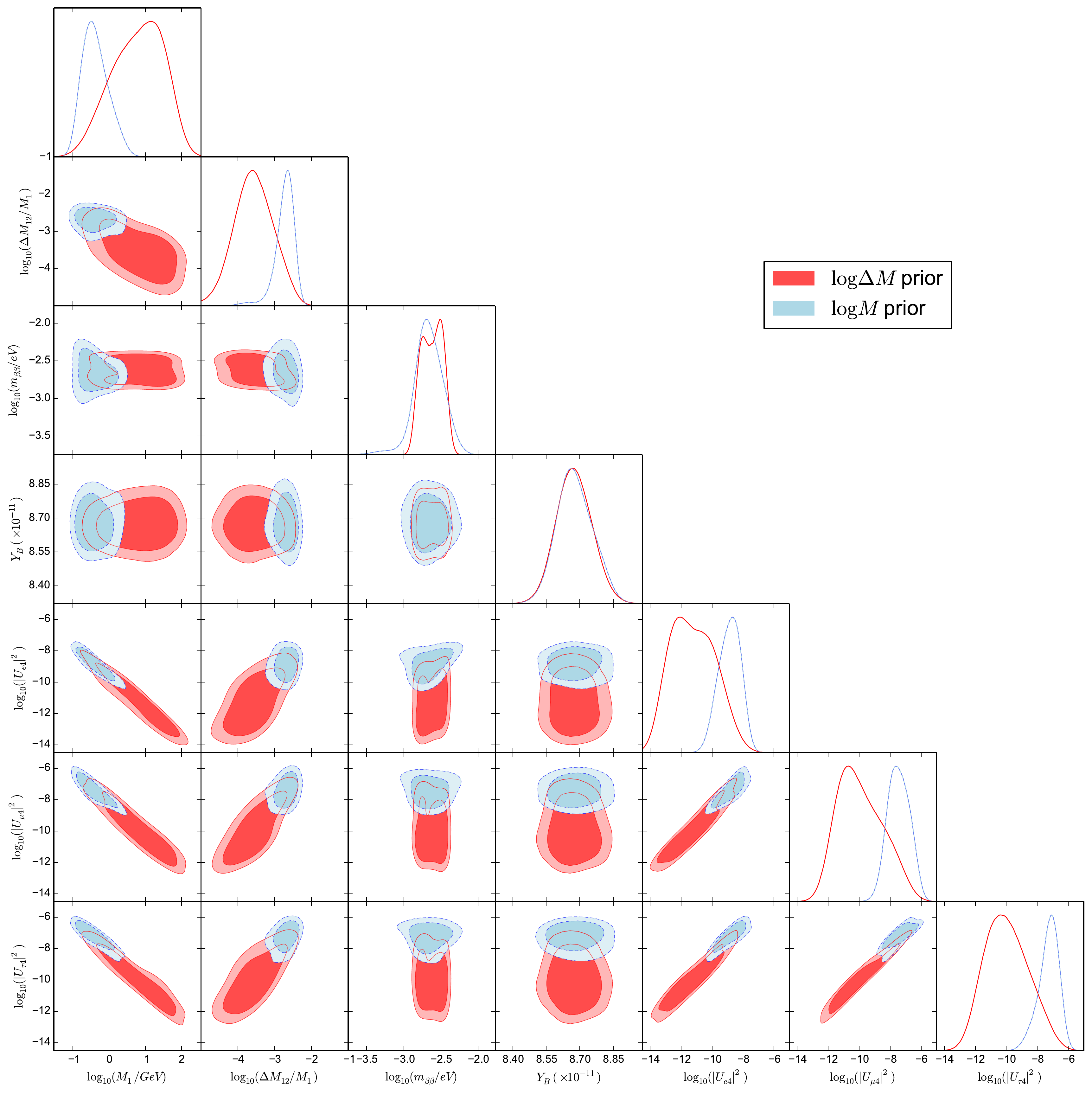} 
\caption{\label{fig:trianglen2nh} Triangle plot with 1D posterior probabilities and  2D $68\%$ and $90\%$ probability contours in the $N=2$ scenario for NH. The parameters shown are the observables $M_1$, $\Delta M_{12}/M_1=(M_2-M_1)/M_1$, $m_{\beta\beta}$, $Y_B$,  and the three mixings with the first of the heavy states $|U_{\alpha 4}|^2$ for $\alpha=e,\mu, \tau$. The blue and red contours correspond respectively to the assumption of a flat prior in $\log_{10} M_1$ and $\log_{10} M_2$ and to a flat prior in $\log_{10} 
M_1$ and $\log_{10}(\Delta M_{12})$.
 }
\end{center}
\end{figure}

\begin{figure}[h]
 \begin{center}
\includegraphics[scale=0.32
]{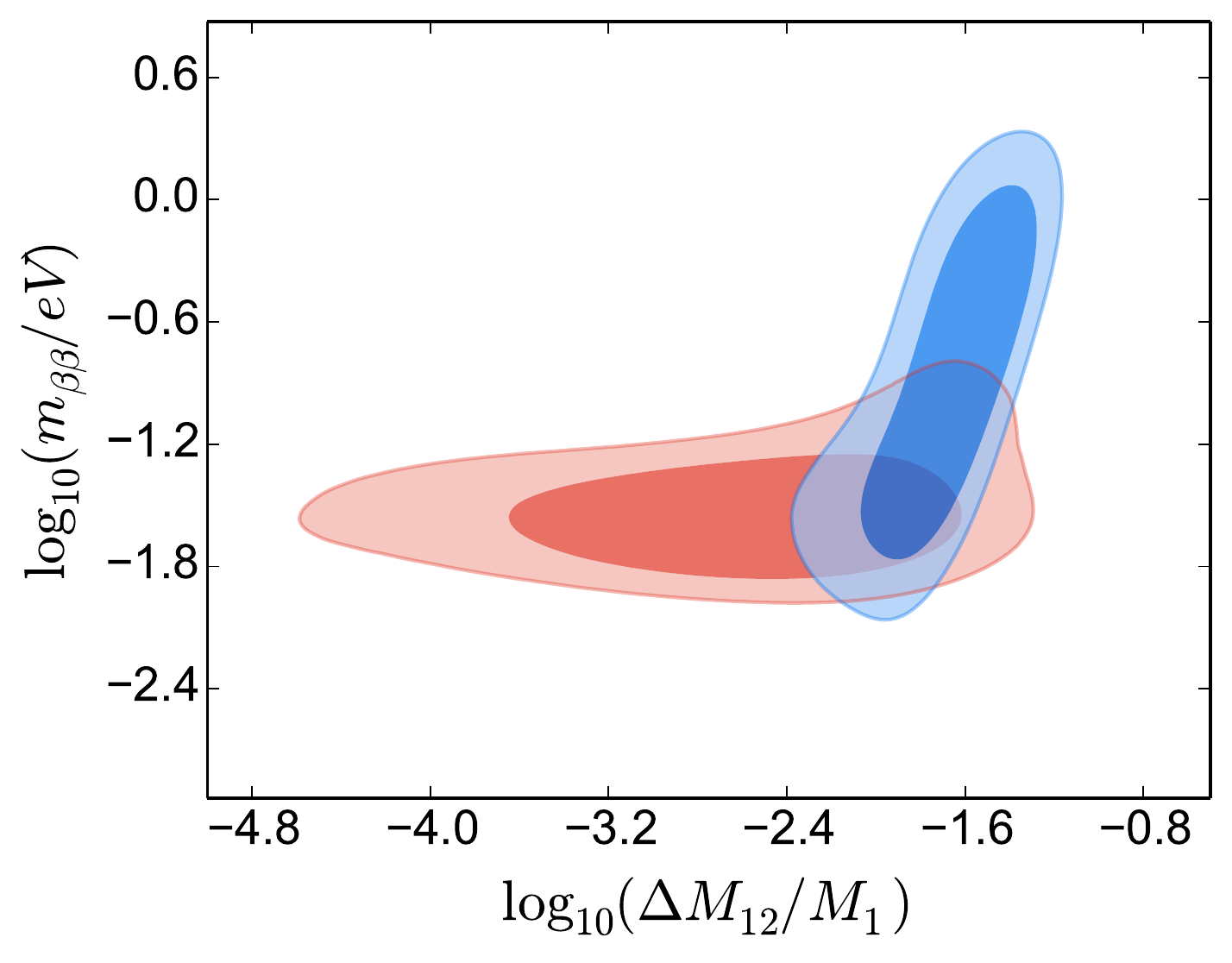} \includegraphics[scale=0.32]{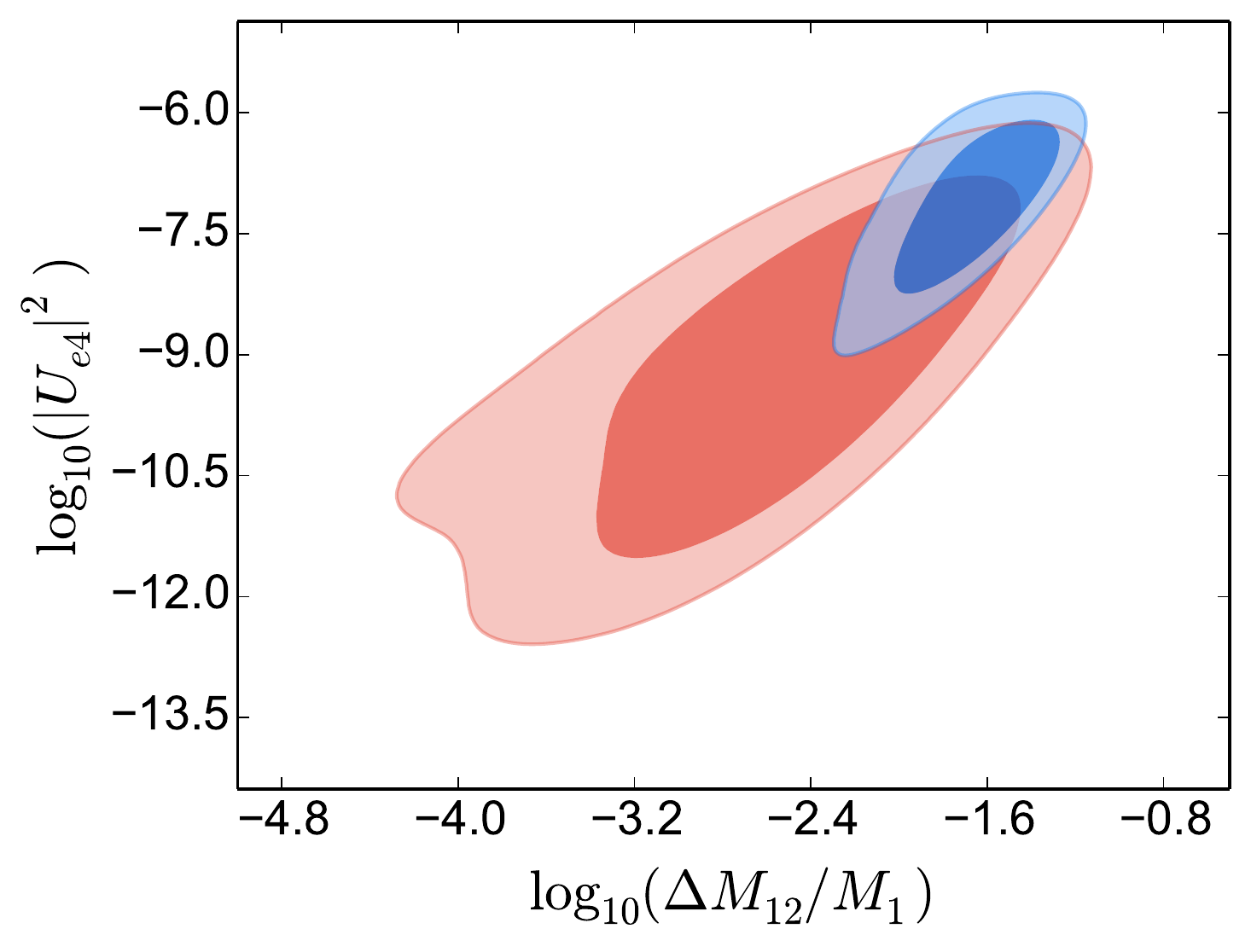} \includegraphics[scale=0.32]{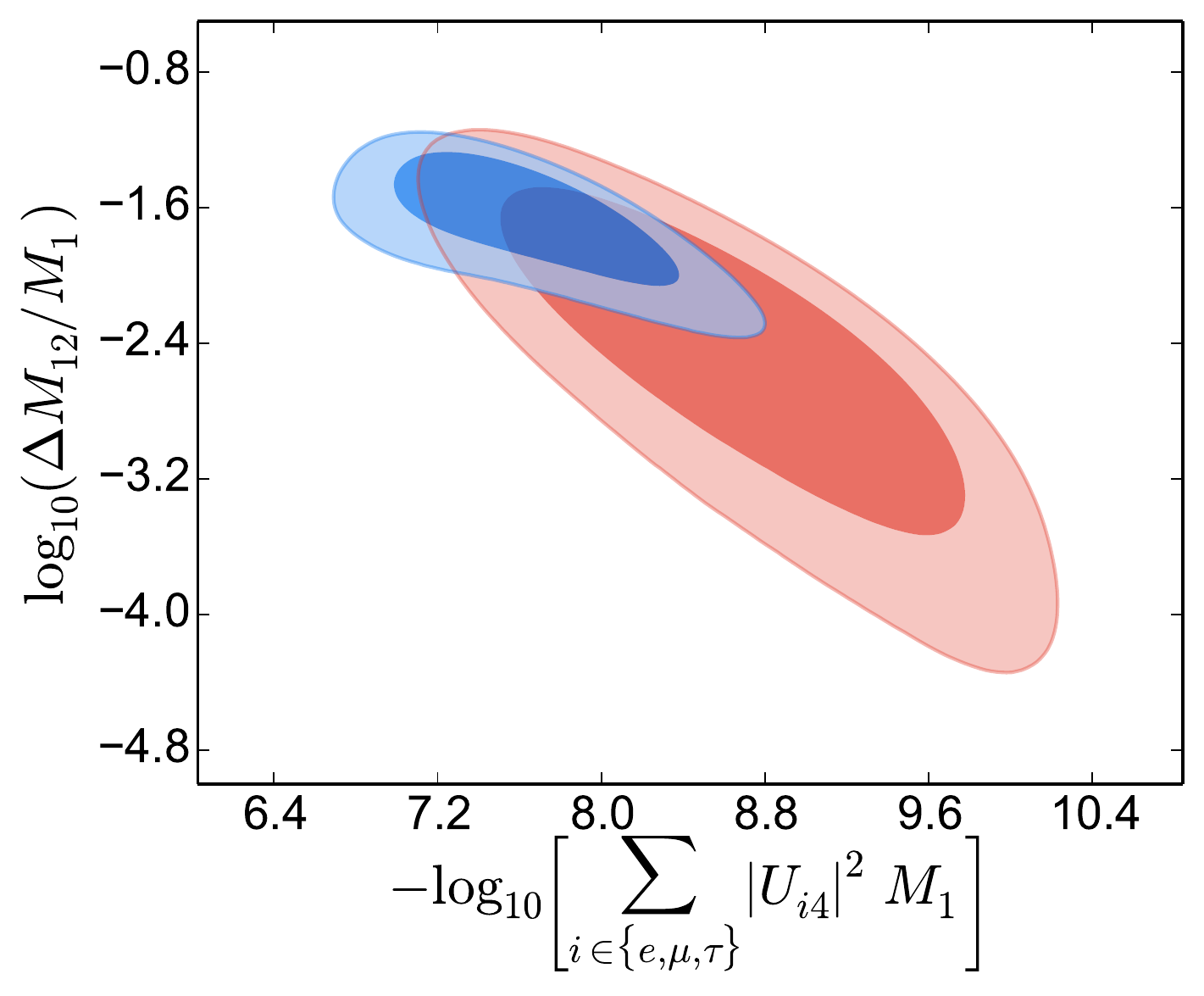} 
\caption{\label{fig:zoom}  Posterior probabilities for the amplitude of neutrinoless double beta decay (left), electron mixing (middle) and $\sum_{\alpha=e,\mu,\tau} |U_{\alpha
 4}|^2 M_1$ (right) versus the mass degeneracy.}
\end{center}
\end{figure}

Figures~\ref{fig:trianglen2ih} and \ref{fig:trianglen2nh} show, for IH and NH, the posterior probabilities of the spectrum of the two relevant states, $M_1, M_2$, the active-sterile mixings of the first heavy state $|U_{\alpha 4}|^2$ (those of the second state are almost identical), the neutrinoless double beta decay effective mass  $|m_{\beta\beta}|$ and the baryon asymmetry $Y_B$. An important consideration are the priors. We have considered flat priors in all the Casas-Ibarra parameters except the masses where we assume a flat prior in $\log_{10}\left({ M_1\over{\rm GeV}}\right)$, within the range $M_1 \in [0.1{\rm GeV} , 10^2 {\rm GeV}]$, and two possibilities: 1) a flat prior also in $\log_{10} \left({M_2\over GeV}\right)$ in the same range or 2) a flat prior in $\log_{10} \left({|M_2-M_1|\over {\rm GeV}}\right)$ in the range $M_2-M_1 \in [10^{-8}{\rm GeV} , 10^2{\rm GeV}]$. The two different colours (light blue and red) in Fig.~\ref{fig:trianglen2ih} correspond to the two options. The significant differences between the two posteriors show the effect of allowing or not for fine-tuning in the degeneracy of the two heavy states. Even though the contours are typically larger if more fine-tuning is allowed, we find interesting solutions with a mild degeneracy, which tend to imply smaller $M_1, M_2$,  larger values of the active-sterile mixing parameters and a sizeable non-standard contribution to neutrinoless double beta decay, which obviously imply much better chances of testability. Figs.~\ref{fig:zoom} zoom in the most interesting results from this study: the mild level of fine-tuning of the blue contours (neither a strong degeneracy is required, nor a very large deviation from the naive seesaw scaling of the mixings), correlated with a relatively large mixing, and a sizeable amplitude for neutrinoless double beta decay. We will come back to this point in the next section. 

In table~\ref{tab:1sigma} we show the 68$\%$ probability ranges for the relevant parameters as extracted from the 1D posterior probabilities, for the two neutrino orderings and the two prior choices. The ranges for the mixings of the second heavy state, $|U_{\alpha 5}|^2$, are basically the same.  

\vspace{0.5cm}
\begin{table}[ht]
\begin{center}
\begin{tabular}{cccccccc}
\hline\hline
NO & Prior & $M_1({\rm GeV})$  & $\Delta M_{12}({\rm GeV})$ & $|U_{e4}|^2$ & $|U_{\mu 4}|^2$ & $|U_{\tau 4}|^2$ & $m_{\beta\beta}({\rm eV})$ \\
\hline
IH & M& $-0.55^{+0.16}_{-0.38}$  & $-2.23^{+0.22}_{-0.19}$  &  $-7.2^{+0.9}_{-0.4}$&$-8.5^{+1.0}_{-0.6}$  & $-8.5^{+0.8}_{-0.7}$ & $-0.84\pm 0.55$\\
 & $\Delta$M & $0.23^{+0.68}_{-0.82}$  & $-2.36^{+0.71}_{-0.51}$  &  $-9.2^{+1.7}_{-1.4}$&$-10.1^{+1.5}_{-1.2}$  & $-9.9^{+1.4}_{-1.2}$ & $-1.48^{+0.15}_{-0.28}$\\
\hline
NH & M &  $-0.39^{+0.31}_{-0.42}$ & $-3.1\pm0.4$ & $-8.9^{+0.8}_{-0.7}$  & $-7.4\pm0.7$  & $-7.3^{+0.7}_{-0.5}$  & $-2.66\pm 0.20$\\
& $\Delta$M & $0.8^{+0.82}_{-0.66}$  & $-2.76\pm 0.62$  & $-11.2^{1.4}_{-1.6}$ & $-9.9^{+1.3}_{-1.8}$  & $-10.0^{+1.3}_{-1.6}$ & $-2.62\pm 0.14$ \\
\hline\hline
\end{tabular}
\caption{For the minimal model $N=2$: $68\%$ posterior probability ranges of $\log_{10}($param) assuming flat prior in
$\log_{10}(M_2({\rm GeV}))$ (M) or $\log_{10}(\Delta M_{12}({\rm GeV}))$ ($\Delta$M). } 
\label{tab:1sigma}
\end{center}
\end{table} 


\begin{figure}[h]
 \begin{center}
\includegraphics[scale=0.32]{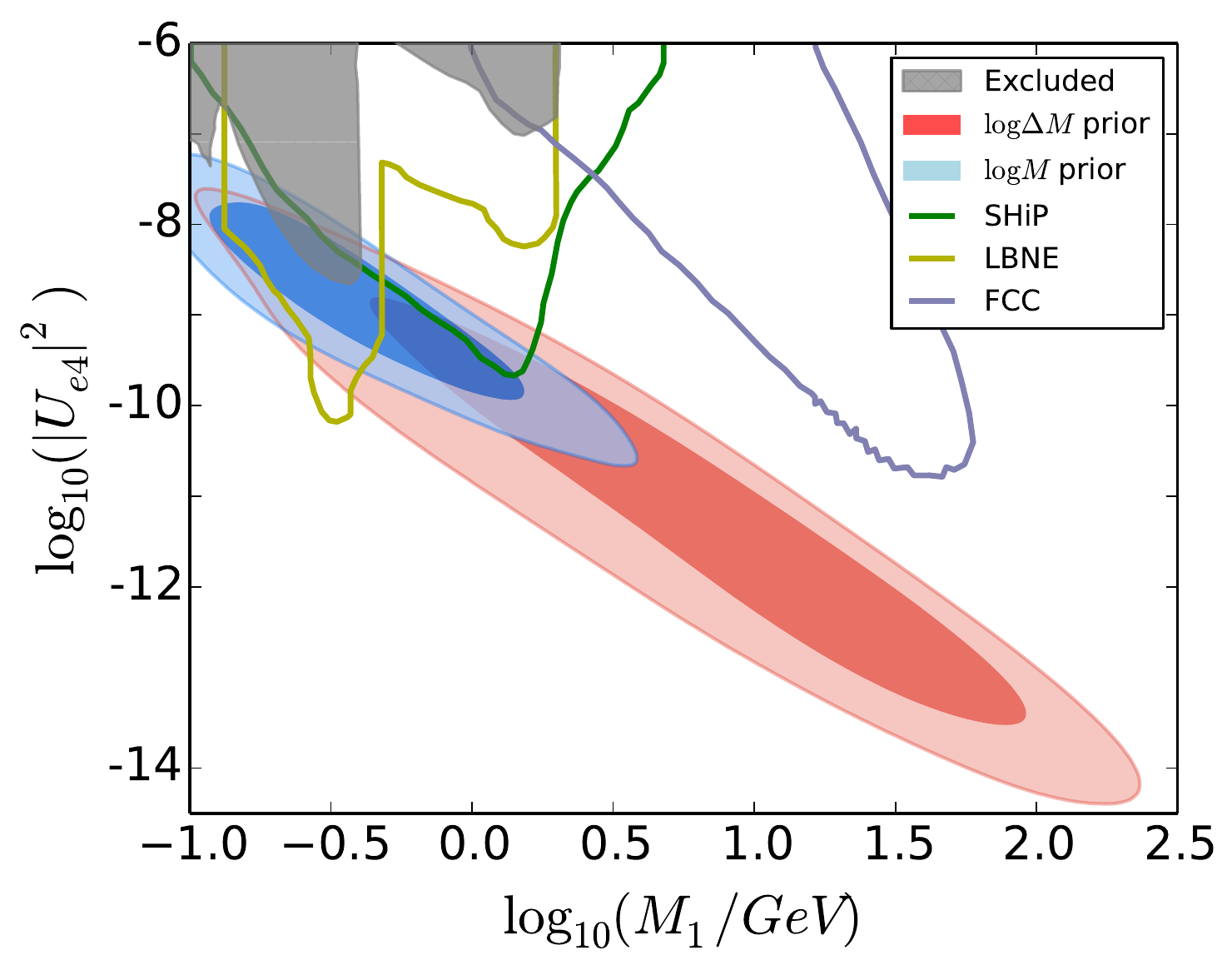} \includegraphics[scale=0.32]{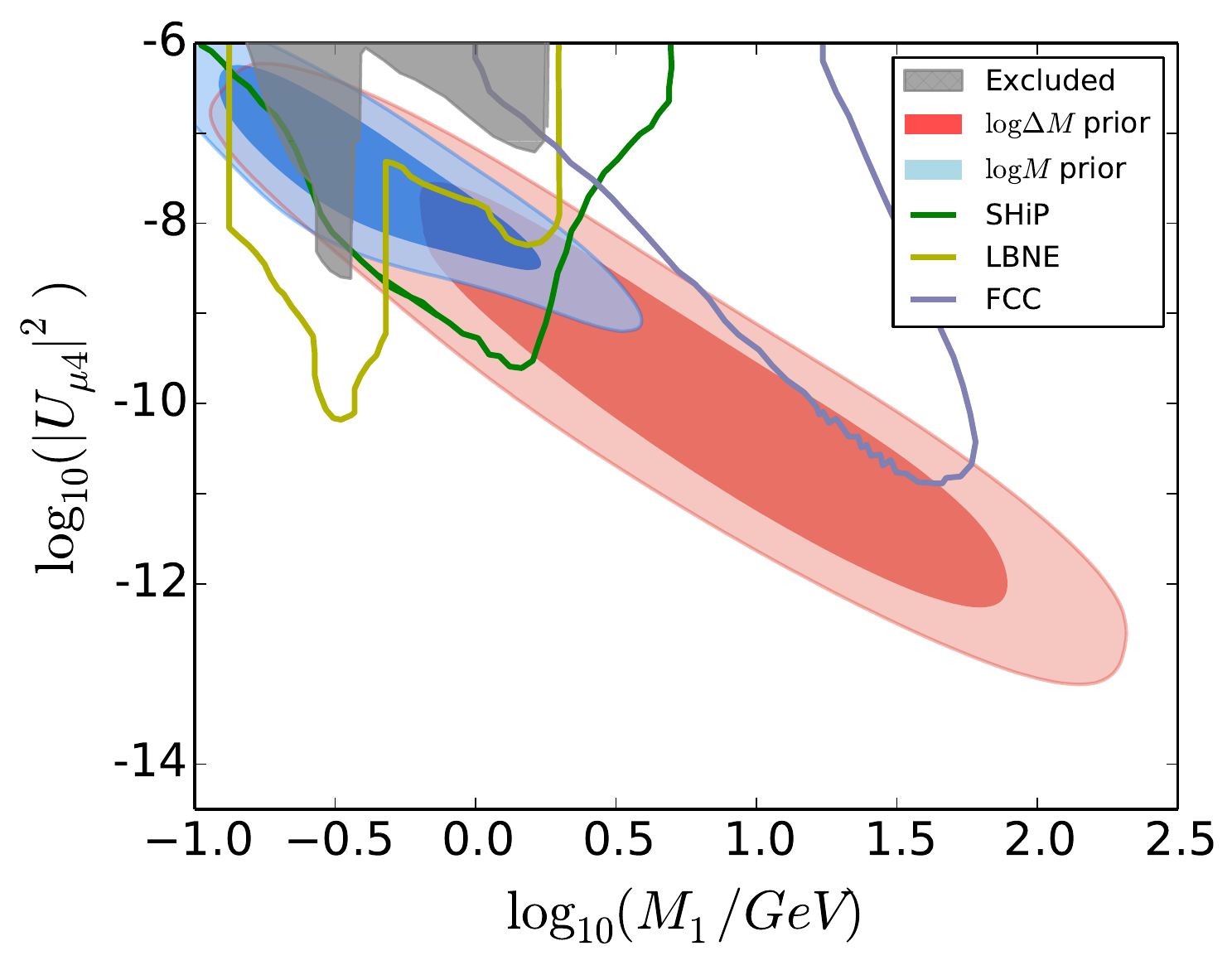} \includegraphics[scale=0.32]{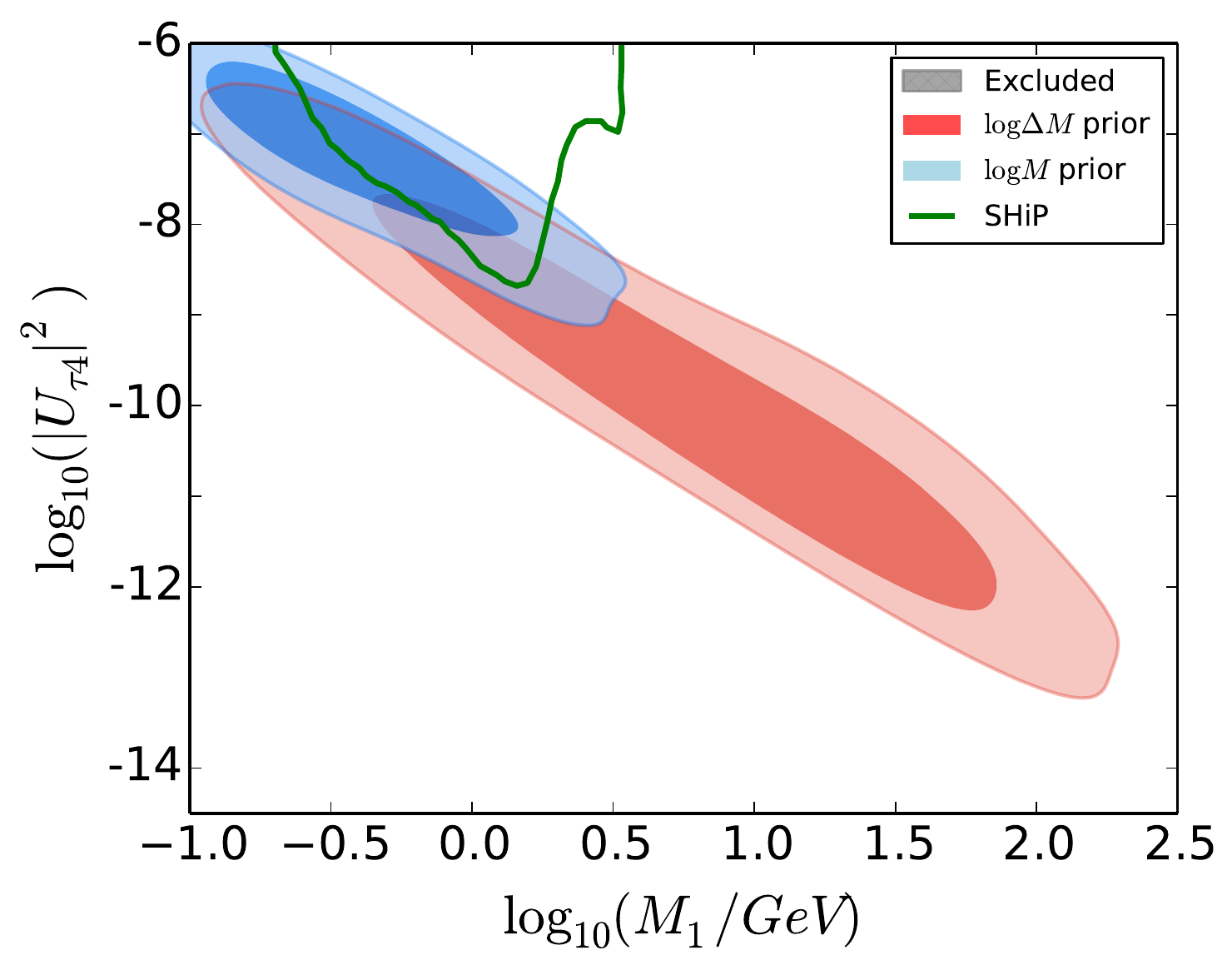} 
\includegraphics[scale=0.32]{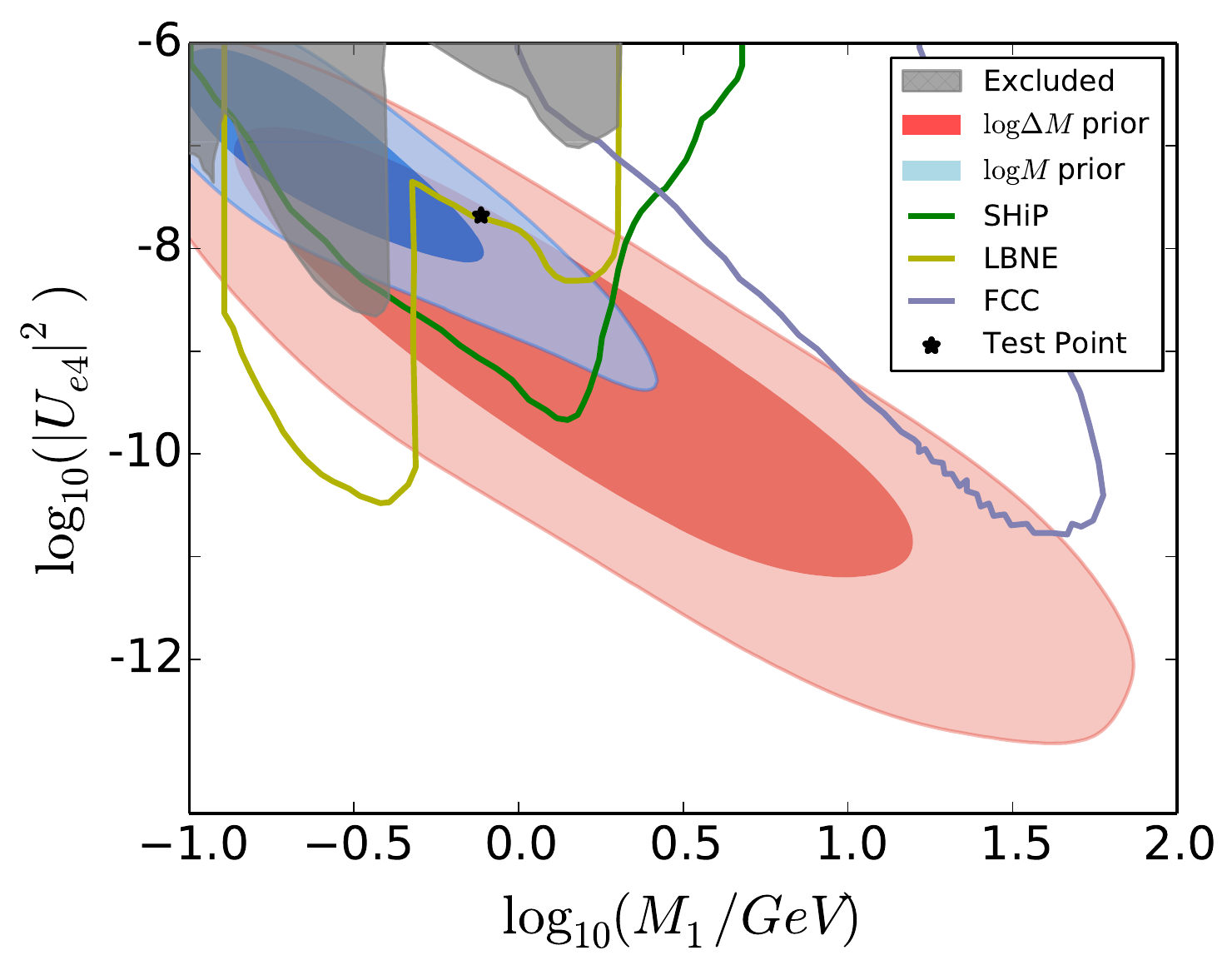} \includegraphics[scale=0.32]{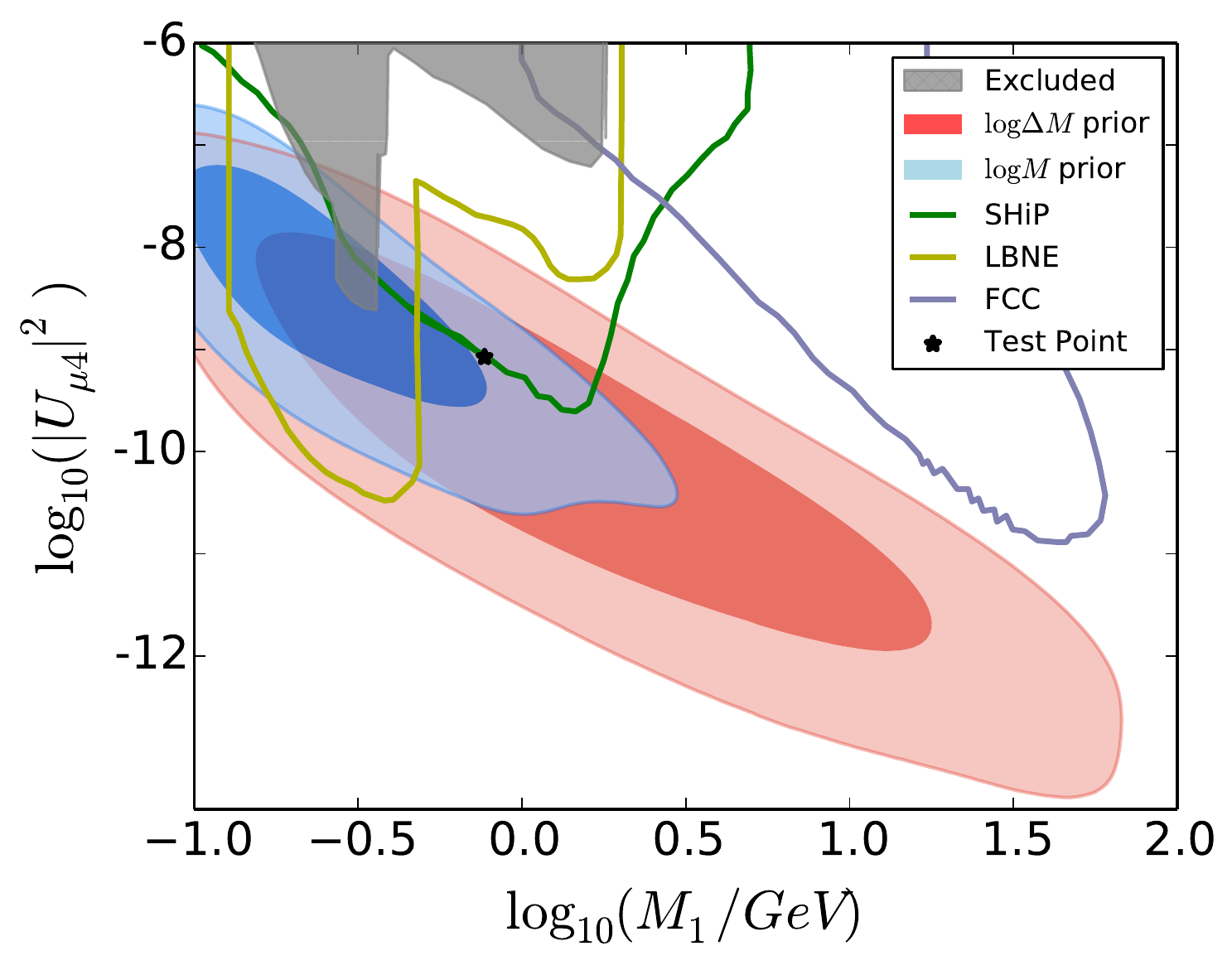} \includegraphics[scale=0.32]{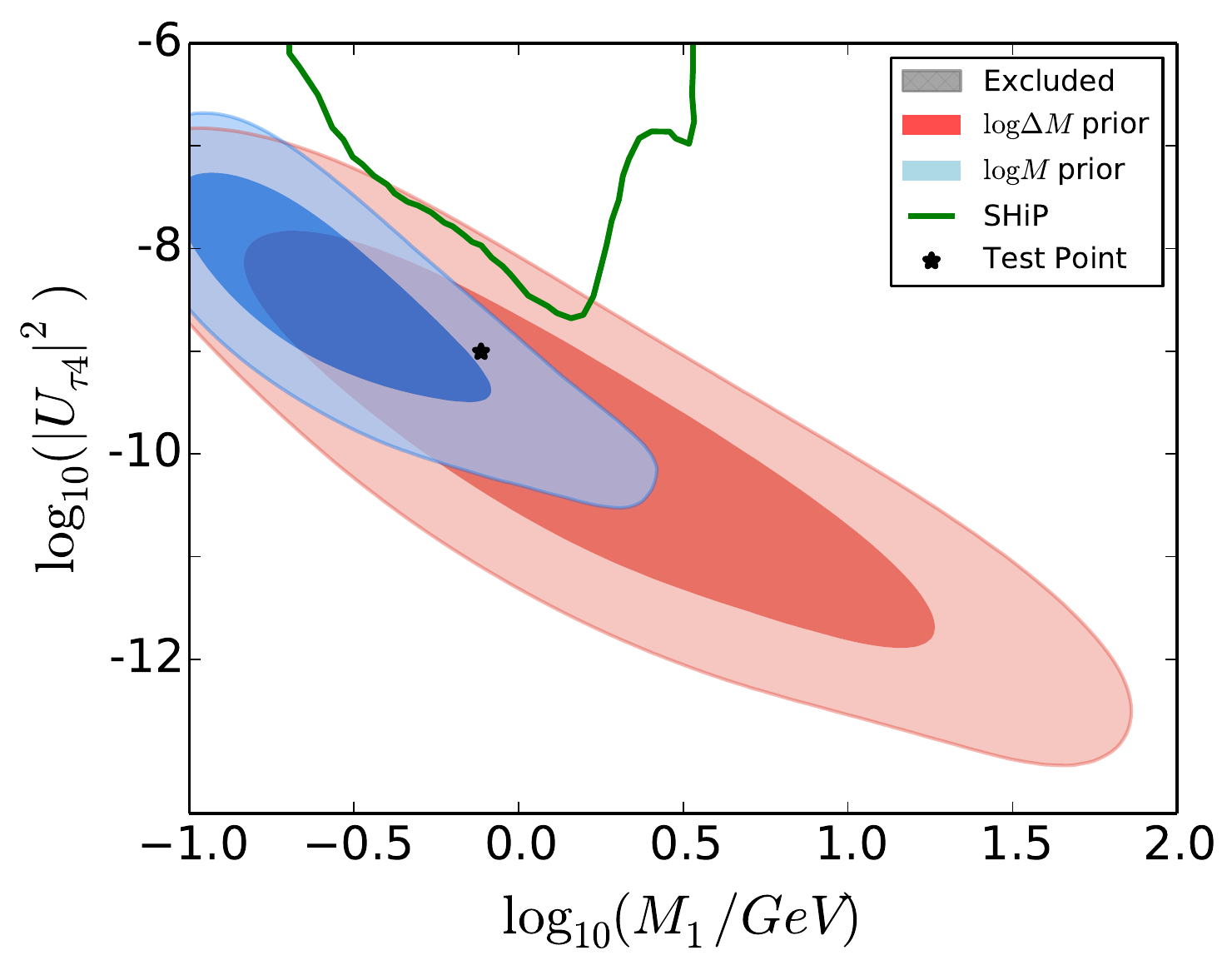} 
\caption{\label{fig:mixings} Comparison of the posterior probability contours at 68$\%$ and 90$\%$ on the planes mixings with $e,\mu,\tau$ versus masses, with the present (shaded region) and future constraints from DUNE, FCC and SHiP for NH (up) y IH (down). }
\end{center}
\end{figure}

 In Figures \ref{fig:mixings} we zoom in the probability plots for the heavy neutrino mixings versus mass and compare them with present \cite{Atre:2009rg},\cite{Deppisch:2015qwa} and future constraints from DUNE \cite{Adams:2013qkq}, SHiP \cite{Alekhin:2015byh} and FCC \cite{Blondel:2014bra}. Constraints from Big Bang Nucleosynthesis  are very restrictive in the low mass range, particularly below the pion threshold \cite{Dolgov:2000jw,Ruchayskiy:2012si}.
In ref.~\cite{Hernandez:2015wna} similar figures were shown from a scan of parameter space assuming only flat priors in $\log_{10} M_1$ and $\log_{10} |M_2-M_1|$. We note that the regions we show here are the result of a full numerical treatment, that avoids any constraint in parameter space and successfully explain the baryon asymmetry within its small $1\%$ uncertainty.  The most important addition is however that of the blue contours that use flat priors in $\log_{10} M_1$ and $\log_{10} M_2$, and therefore avoid too large fine-tuning. These solutions point to 
a region of parameter space within SHiP reach as the most probable one.
It is interesting that the sensitivity of SHiP and DUNE to the $e$ or $\mu$ channels will cover to a large extent the blue regions.
 When a larger degree of degeneracy in the masses is allowed (red regions), the right baryon asymmetry can be obtained also for larger masses, up to 10-100 GeV, but this high mass region will be harder to test experimentally (for recent work see also \cite{Antusch:2016vyf}). 

\vspace{0.5cm}
In the $N=3$ case, there are 13 unknown parameters and the exploration of parameter space is significantly more challenging. This case will be considered elsewhere.

\section{Predicting the baryon asymmetry in the minimal $N=2$ model}

A very relevant question is whether the baryon asymmetry could be predicted in this scenario if the heavy sterile neutrinos are within reach of future experiments, such as the SHiP experiment. We will analyse this question in the simplest case $N=2$ where the number of unknown parameters is minimal. Obviously the situation for the $N=3$ case will be much more difficult. 

The SHiP experiment will be capable of detecting heavy neutrinos in the few GeV range provided their mixings are sufficiently large. In particular  significantly larger
than what the naive seesaw scaling  $|U_{ai}|^2 \sim m_i/M_i$ would suggest. In the Casas-Ibarra parametrization of eq.~(\ref{eq:yci}), this implies that the entries of the $R$ matrix 
need to be significantly larger than one, and therefore the imaginary part of the complex angle needs to be sizeable.  

In order to understand the dependence of $Y_B$ on the different  parameters, it is useful to consider the perturbative results 
of \cite{Hernandez:2015wna}.  The CP asymmetries responsible for the baryon number generation, $\Delta_{CP}$, in the weak washout regime, can be generically written as 
\be 
 \Delta_{CP}= \sum_{\alpha, k} |Y_{\alpha k} |^2 \,  \Delta_\alpha,
 \label{DeltaCP1}
\ee
with
\be 
\label{DeltaCP2}
\Delta_\alpha =
\sum_{i,j}  {\rm Im} [ Y_{\alpha i}  Y^*_{\alpha j} (Y^\dagger Y)_{ij} ] f(M_i,M_j).
\ee
For the $N=2$ case, when $\phi_{i3}=0$ and $y_3=0$,  this quantity can be written as \cite{Hernandez:2015wna} 
\be
Y_B = \Delta_{CP} = y_1 y_2 (y_2^2-y_1^2) \left( (y_2^2 -y_1^2) I_1^{(2)} + y_2^2  I_1^{(3)}\right)g(M_2,M_1),
\ee
where the invariants $I_1^{(2),(3)}$ are defined in eq.~(\ref{eq:cpinvs}).
This is indeed the dependence obtained from the solution of the kinetic equations in the perturbative approximation obtained in \cite{Hernandez:2015wna}, in the weak washout limit. For our new kinetic equations, the perturbative result is shown in eq.~(\ref{eq:anal}) (where only the $I_1^{(2)}$ contribution has been kept).  We can read from eq.~(\ref{eq:anal}) in the limit of weak washout,  $\Gamma_\pm t \ll 1$, and using eq.~(\ref{eq:yb}):
\be
g(M_2,M_1) = 1.3 \times 10^{-3} \times 2 \left({2 \over 3}\right)^{4\over 3} {\pi^{3/2} \over \Gamma(-1/6)}  \left({9 \xi(3)\over \pi^2}\right)^2 {(\gamma^{(0)})^2 \gamma^{(1)} \over M_P^{*2/3}  T_{EW}}  \frac{{\rm sign}(\Delta M_{12}^2)}{|\Delta M_{12}^2|^{2/3}}.
\ee
In contrast with eq.~(\ref{eq:anal}), this approximation fails at large times since it is valid only for $\Gamma_\pm t \ll 1$. It typically overestimates the asymmetry, but should give qualitatively the right dependence on the parameters. 

What we need however are the expressions in terms of the Casas-Ibarra parameters. The relations are typically very complicated, but we can identify a few small parameters and perturb in them:
\be
\mathcal{O}\left(\epsilon \right): r\equiv\sqrt{\frac{\Delta m^2_{sol}}{\Delta m^2_{atm}}}\sim  \theta_{13}  \sim e^{-{\gamma\over 2}},
\ee
where $\gamma$, assumed positive, is the imaginary part of the complex angle of the $R$ matrix that needs to be large to avoid the naive seesaw scaling of the active-sterile mixings \footnote{
Note that $\gamma$ can also be negative, but there is an approximate symmetry $\gamma\rightarrow -\gamma$, that would lead to very similar results by expanding in $e^{-{|\gamma|\over 2}}$ in this case. }.
Defining 
\begin{eqnarray}
A\equiv \frac{e^{2\gamma}\sqrt{\Delta m^2_{atm}}}{4},
\end{eqnarray}
the result for the heavy-light mixing for IH can be written as
\begin{eqnarray}
|U_{e4}|^2 M_1 \simeq |U_{e5}|^2 M_2 &\simeq& A \Big[ (1 + \sin\phi_1\sin2\theta_{12})(1-\theta_{13}^2)  + {1\over 2} r^2 s_{12} (c_{12} \sin\phi_1+s_{12})+{\mathcal O}(\epsilon^3)\Big] ,\nonumber\\
|U_{\mu4}|^2 M_1 \simeq |U_{\mu5}|^2 M_2 &\simeq&A \Big[ \left
(1 - \sin\phi_1\sin2\theta_{12} \left(1+ {1\over 4}r^2 \right)+ {1\over 2} r^2 c^2_{12} \right) c_{23}^2 \nonumber\\
&&+ \theta_{13} (\cos \phi_1 \sin \delta - \sin\phi_1 \cos 2 \theta_{12} \cos\delta) \sin 2 \theta_{23}\nonumber\\
&&+ \theta_{13}^2 (1+\sin\phi_1 \sin 2 \theta_{12}) s_{23}^2  +{\mathcal O}(\epsilon^3)\Big],\nonumber\\
|U_{\tau4}|^2 M_1 \simeq |U_{\tau5}|^2 M_2 &\simeq&A \Big[ \left
(1 - \sin\phi_1\sin2\theta_{12} \left(1+ {1\over 4} r^2\right)+ {1\over 2} r^2 c^2_{12} \right) s_{23}^2  \nonumber\\
&&- \theta_{13} (\cos \phi_1 \sin \delta - \sin\phi_1 \cos 2 \theta_{12} \cos\delta) \sin 2 \theta_{23}\nonumber\\
&&+ \theta_{13}^2 (1+\sin\phi_1 \sin 2 \theta_{12}) c_{23}^2  +{\mathcal O}(\epsilon^3)\Big].\nonumber\\
\label{eq:uihanal}
\end{eqnarray}
The result for NH is:
\begin{eqnarray}
|U_{e4}|^2 M_1 \simeq |U_{e5}|^2 M_2 &\simeq& A\Big[  r s_{12}^2  -2  \sqrt{r} \theta_{13} \sin(\delta+\phi_1) s
_{12}+ \theta_{13}^2 +{\mathcal O}(\epsilon^{5/2})\Big] ,\nonumber\\
|U_{\mu4}|^2 M_1 \simeq |U_{\mu5}|^2 M_2 &\simeq&A\Big[ s_{23}^2  - \sqrt{r}~  c_{12} \sin\phi_1 \sin 2\theta_{23}  + r c_{12}^2 c_{23}^2 \nonumber\\
&&+ 2  \sqrt{r} ~\theta_{13} \sin(\phi_1+\delta) s_{12} s_{23}^2 - \theta_{13}^2 s_{23}^2 +{\mathcal O}(\epsilon^{5/2})\Big],\nonumber\\
|U_{\tau4}|^2 M_1 \simeq |U_{\tau5}|^2 M_2 &\simeq& A \Big[ c_{23}^2  + \sqrt{r}  c_{12} \sin\phi_1 \sin 2\theta_{23}  + r c_{12}^2 s_{23}^2\nonumber\\
&&+ 2  \sqrt{r} ~\theta_{13}\sin(\phi_1+\delta) s_{12} c_{23}^2 -\theta_{13}^2 c_{23}^2 +{\mathcal O}(\epsilon^{5/2})\Big].\nonumber\\
\label{eq:unhanal}
\end{eqnarray}

Note that the mixings  depend exponentially on $\gamma$ and are inversely proportional to the heavy masses, but this dependence drops in any ratio. 

At leading order (LO) in the $\epsilon$ expansion, we see that for IH, the ratio of  the electron and muon mixings is independent of $\gamma$ and heavy masses and depends exclusively on  the Majorana CP phase of the PMNS matrix, $\phi_1$:
\begin{eqnarray}
{\rm IH}: ~~{|U_{e4}|^2 \over |U_{\mu 4}|^2} \simeq {1 \over c_{23}^2}{1 + \sin\phi_1\sin2\theta_{12}\over 1 - \sin\phi_1\sin2\theta_{12} } +{ \mathcal O}(\epsilon).
\label{eq:lorat}
\end{eqnarray}
 This is important because the CP asymmetry strongly depends on this phase as we will see below, therefore for the IH, the putative measurement of the masses and the mixings of these states in  SHiP for example, would allow in principle to fix simultaneously $\gamma$ and $\phi_1$. For NH on the other hand the SHiP measurement would only provide information on $\gamma$ but not on $\phi_1$ nor any other phase, at this order:
\begin{eqnarray}
{\rm NH}: ~~{|U_{e4}|^2 \over |U_{\mu 4}|^2} \simeq 2 r { s_{12}^2 \over s_{23}^2}+{ \mathcal O}(\epsilon^{3/2}).
\end{eqnarray} 
The mixings to taus are at this order the same as those to muons. 

The leading order however is not precise enough. For IH, it is clear from eq.~(\ref{eq:lorat}) that depending on the value of $\phi_1$ a significant suppression of the leading terms in the numerator or denominator can take place, therefore at this point the NLO corrections are relevant and these bring a new undetermined parameter, $\delta$,  as can be seen from eqs.~(\ref{eq:uihanal}), which is the CP phase that can be measured in neutrino oscillations! In this case, the measurement of the ratio of mixings at SHiP cannot resolve $\phi_1$ and $\delta$ simultaneously and a degeneracy between these two phases remains. We will come back to this interesting observation in the following section. Clearly the $\delta$ phase could be determined in future neutrino oscillation experiments. 

At leading order in the $\epsilon$ expansion, the CP asymmetry in this regime can be approximated by

\bea
\left|{\Delta_{CP}\over g(M_1,M_2)}\right|_{IH}&=&e^{4\gamma}\,\frac{(\Delta m^2_{atm})^{3/2}}{4v^6} M_1M_2(M_1+M_2)  \Big[( \sin 2\theta \cos2\theta_{12}-\cos\phi_1 \cos 2\theta\sin2\theta_{12} )\times \nonumber\\
& & \left(\sin^22\theta_{23}+(4 + \cos 4 \theta_{23})\sin\phi_1\sin2\theta_{12}\right)+ {\mathcal O}(\epsilon)\Big],\nonumber\\
\left|\Delta_{CP}\over g(M_1,M_2)\right|_{NH}&=&e^{4\gamma}\,\frac{\, (\Delta m^2_{atm})^{3/2}}{4v^6} \, M_1 M_2(M_1+M_2)\Big[{\sqrt{r}\over 2} \sin4\theta_{23} c_{12}\cos(\phi_1-2\theta)
\nonumber\\
&+&
 \,r \Big(\sin^22\theta_{23}\left[c_{12}^2\sin2(\phi_1-\theta)+(2+\cos2\theta_{12})\sin2\theta\right]-2 \Big)
\nonumber\\
&+&
\sqrt{r}\,\theta_{13}\,s_{12}(1 + \cos^22\theta_{23})\cos(\delta +\phi_1- 2 \theta)
 +{\mathcal O}(\epsilon^2)\Big],
\nonumber\\
\eea
where in the case of NH we have included NLO corrections because the LO cancel for maximal atmospheric mixing. 
We see that for both neutrino orderings, it depends strongly on the real part of the Casas-Ibarra angle $\theta$, which does not affect any other of the observables above. In particular, independently of  the value of $\phi_1$, there is always a value of $\theta$ that can make the asymmetry vanish. For instance for the IH result the value can be approximated by 
\begin{eqnarray}
{\rm IH:} &&~~\tan 2 \theta \simeq \cos \phi_1 \tan 2 \theta_{12}, 
\end{eqnarray}
therefore the uncertainty in $\theta$ forbids to set  a lower bound on the asymmetry, although an upper bound can be derived. 
Therefore even if the sterile states would be discovered at SHiP and their mixings to electrons and muons accurately measured, the asymmetry can only be predicted up to this angle.  

Furthermore as we have seen in order to explain the baryon asymmetry in the $N=2$ case, a significant level of degeneracy between the two states
is needed. The dependence on this quantity enters in  the function $g(M_1,M_2)$. Although we have not found a detailed study of the expected uncertainty in the invariant mass at SHiP, given the momentum resolution for muons and pions quoted in \cite{Anelli:2015pba}, we expect that this uncertainty could be better than 1$\%$.   If the degeneracy cannot be measured, a large uncertainty in the CP asymmetry will  result also from this unknown. 

Interestingly neutrinoless double beta decay also depends on both unknowns: $\theta$ and $\Delta M_{12}$. The effective neutrino mass in neutrinoless double beta is given approximately by \cite{Ibarra:2011xn,Lopez-Pavon:2015cga}
\begin{eqnarray}
|m_{\beta\beta}|_{IH}&\simeq& \sqrt{\Delta m^2_{atm}} \left|c_{13}^2\left(c_{12}^2 + e^{2i \phi_1 }s_{12}^2 \left(1+ {r^2\over 2}\right)
\right)\right.\label{eq:ihmbb}
\\
&-& \left.f(A)\,e^{2i\theta }e^{2\gamma}(c_{12}-i e^{i\phi_1}s_{12})^2 (1-2 e^{i \delta} s_{23} \theta_{13})
\frac{\left(0.9\,\text{GeV}\right)^2}{ 4 M_1^2}
\left(1-\left(\frac{M_1}{M_1+\Delta M_{12}}\right)^2\right)\right|,\nonumber
\\
|m_{\beta\beta}|_{NH}&\simeq&\sqrt{\Delta m^2_{atm}} \left|  e^{2 i \phi_1}
c_{13}^2 s_{12}^2 r +e^{-2 i \delta}
s_{13}^2 \right.
\\
&-&\left.f(A)\,e^{2i \theta}e^{2\gamma}s_{12}\,(r s_{12} e^{2 i \phi_1}-2 i \sqrt{r} \theta_{13}  e^{-i \delta}) 
\frac{\left(0.9\,\text{GeV}\right)^2}{ 4 M_1^2}
\left(1-\left(\frac{M_1}{M_1+\Delta M_{12}}\right)^2\right)\right|,
\nonumber
\end{eqnarray}
where  the two lines in each amplitude correspond respectively to the light and heavy contributions. $f(A)$ depends on the nucleus under consideration. For $^{48}$Ca, 
$^{76}$Ge, $^{82}$Se, $^{130}$Te and $^{136}$Xe, $f(A)\approx$ 0.035, 0.028, 0.028, 0.033 and 0.032, 
respectively \cite{Blennow:2010th,Ibarra:2010xw}. Since $f(A)$ is very small we have neglected ${\mathcal O}(\epsilon^2)$ effects in the heavy contribution. 

Note that the non-standard contribution from the heavy states is very sensitive to the mass degeneracy. Furthermore the interference between the light and heavy contributions  is very sensitive to the angle $\theta$, and it is precisely in the region around 1~GeV where  the heavy and light contributions can be comparable, and 
can effectively interfere \cite{Mitra:2011qr,LopezPavon:2012zg,Lopez-Pavon:2015cga}.  There is therefore the possibility that neutrinoless double beta decay could provide the missing information to predict the baryon asymmetry. 
 
 \vspace{0.5cm}
 
 \begin{figure}[h]
 \begin{center}
\includegraphics[scale=0.55]{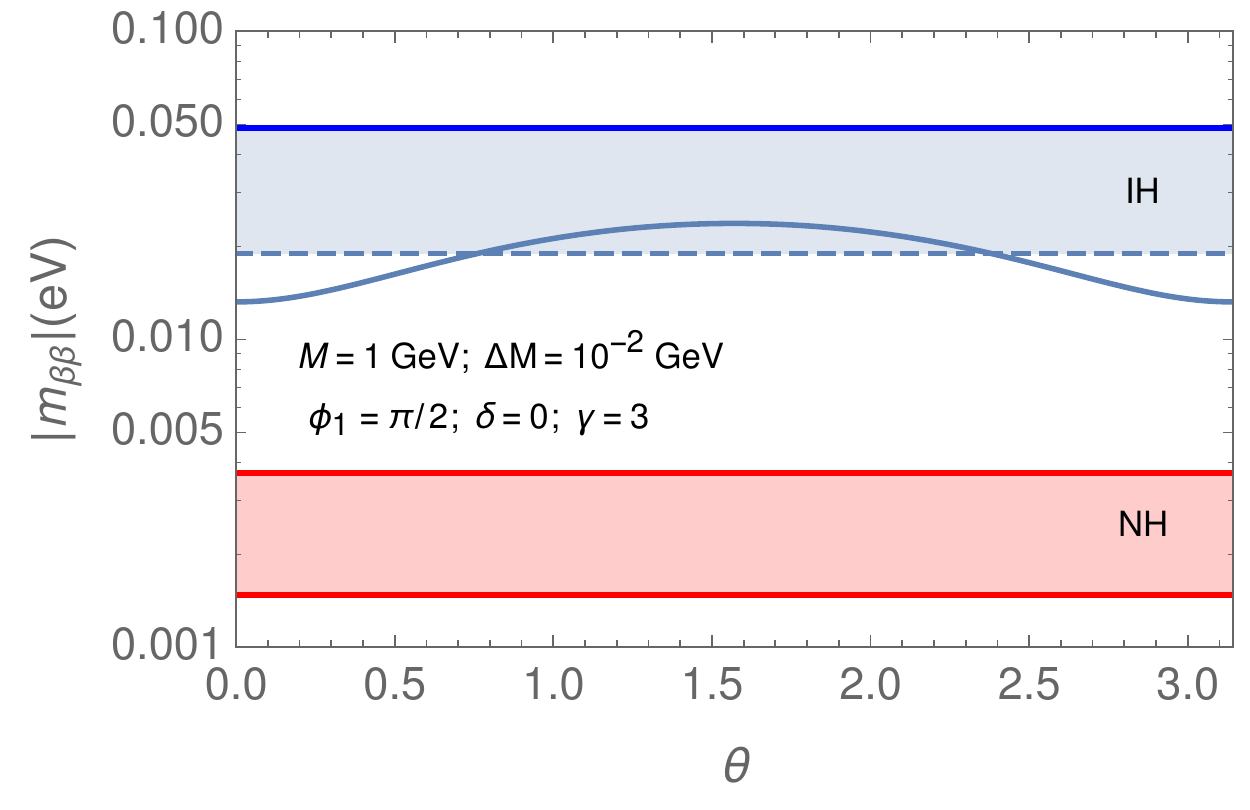} \includegraphics[scale=0.6
]{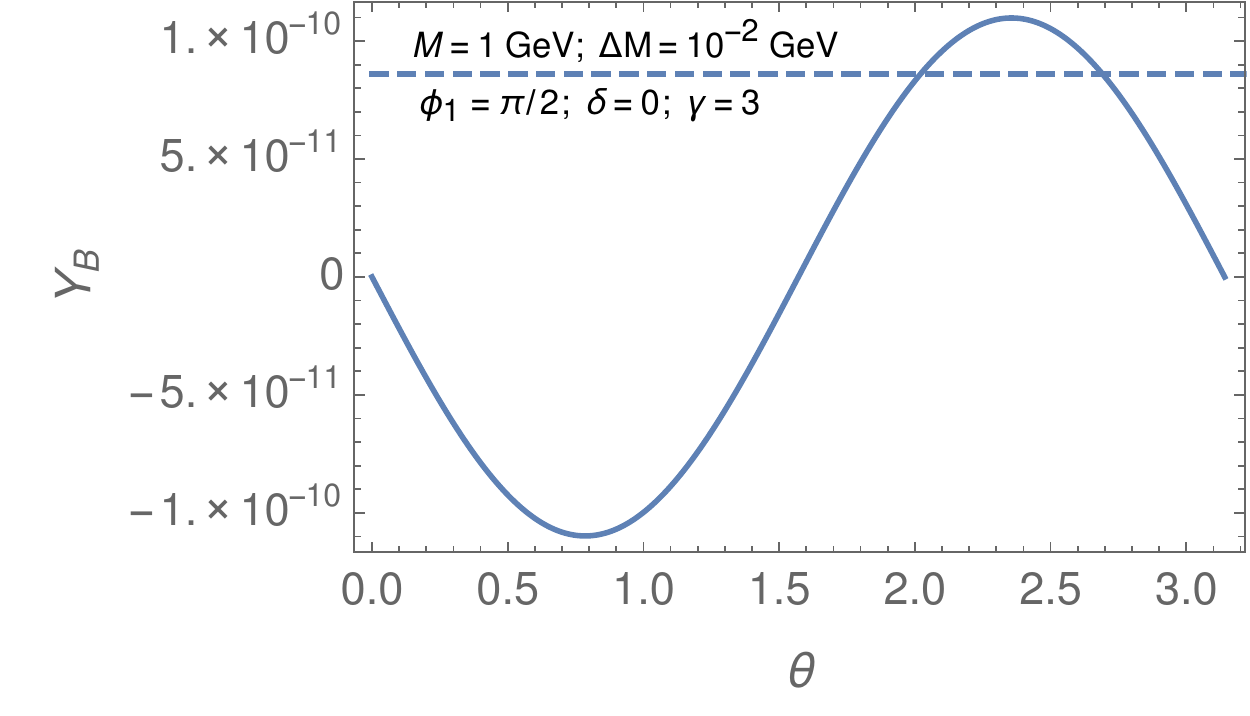} 
\caption{\label{fig:vsth} Left: Dependence of  $m_{\beta\beta}$ on $\theta$ for IH and $M_1=1$GeV$, \Delta M_{12} = 10^{-2}$ GeV, $\phi_1 = {\pi \over 2}$, $\delta=0$ and $\gamma=3$. The red band is the standard $3\nu$ expectation for NH and the blue one that for the IH. The dashed line would be the standard 3$\nu$ contribution for the chosen value of $\phi_1$. Right: $Y_B$ versus $\theta$ for the same parameters. The dashed line is the experimental value of $Y_B$.}
\end{center}
\end{figure}

\begin{figure}[h]
 \begin{center}
\includegraphics[scale=0.85]{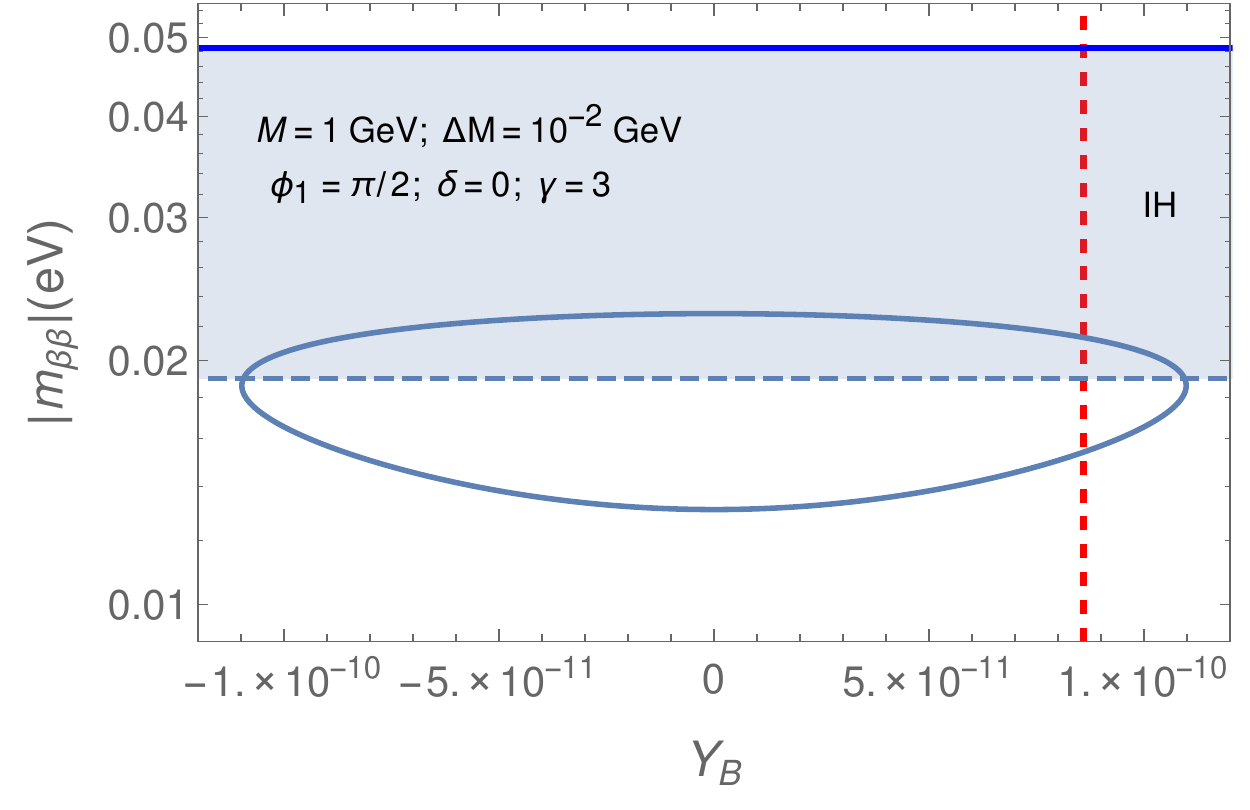}
\caption{\label{fig:mbbvsYB} Correlation of $|m_{\beta\beta}|$ and $Y_B$ when the parameters that could in principle be measured at SHiP 
are fixed and the neutrino ordering is inverted. The  band is the standard $3\nu$ expectation. The vertical dashed line is the measured $Y_B$, and 
the horizontal one corresponds to the $3 \nu$ expectation for $\phi_1 = {\pi \over 2}$.}
\end{center}
\end{figure}

On the left plot of Fig.~\ref{fig:vsth} we show $|m_{\beta\beta}|$ as a function of the angle $\theta$ for IH and some choice of parameters that are within the range of SHiP and assumed known. $|m_{\beta\beta}|$ has been computed exactly using the nuclear matrix elements for $^{76}$Ge from ref.~\cite{Blennow:2010th}.The bands are the standard $3\nu$ contributions to $|m_{\beta\beta}|$ for NH/IH. If the uncertainty in $|m_{\beta\beta}|$ could be better than the spread within the standard IH region, a determination of $\theta$ could result from this measurement. On the right plot  we show the dependence of $Y_B$  (computed exactly) on the same angle. In Fig.~\ref{fig:mbbvsYB} we show the correlation between $|m_{\beta\beta}|$ and $Y_B$ as we vary $\theta$. 
Ideally a precise determination of $|m_{\beta\beta}|$ could predict the baryon asymmetry up to a global sign. In practice, this would require 
a very good control of the nuclear matrix elements which is very challenging.

\begin{figure}[h]
 \begin{center}

\includegraphics[scale=0.6]{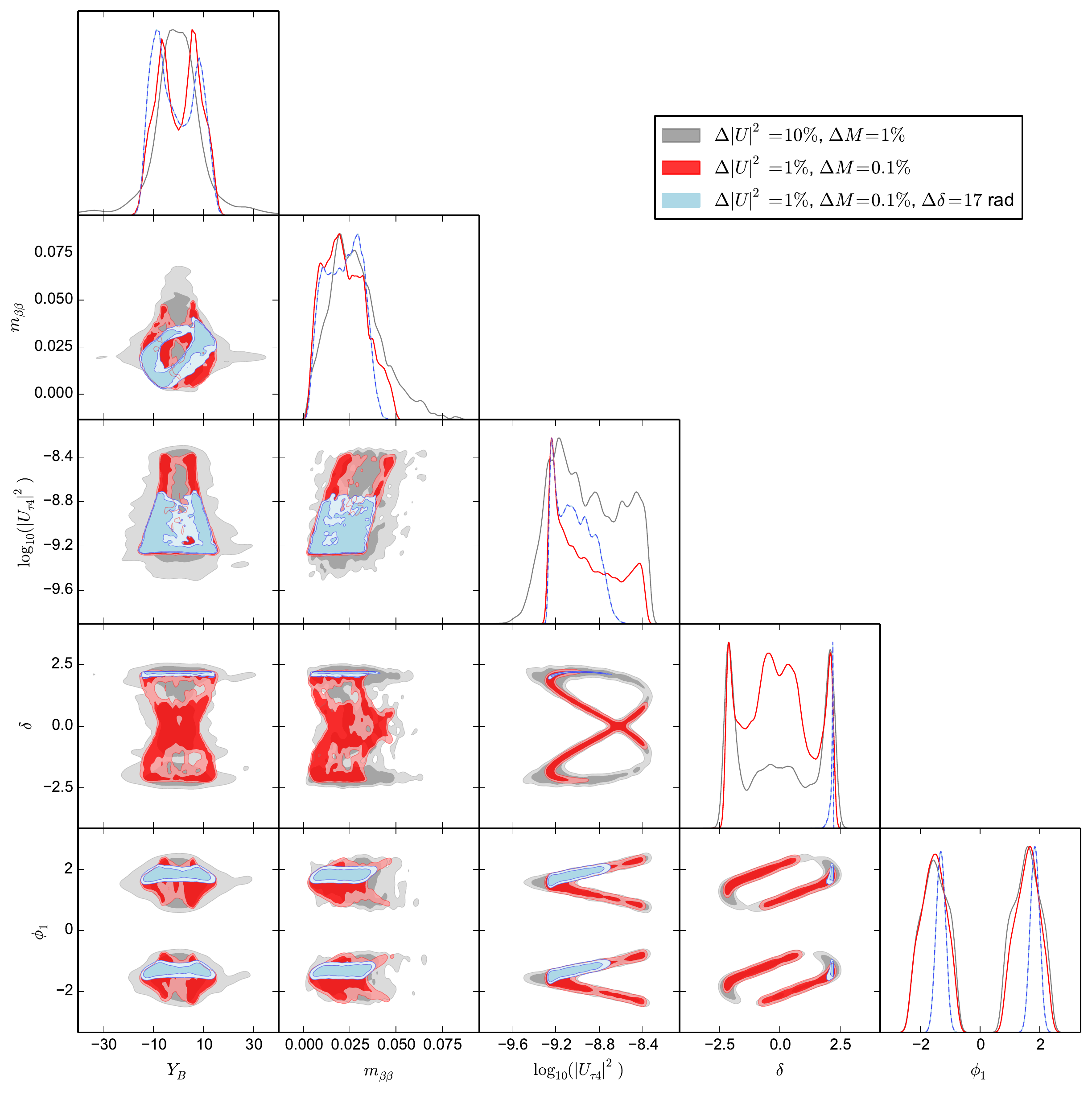} 
\caption{\label{fig:triangleship} Triangle plot with 1D posterior
  probabilities and  2D $68\%$ and $90\%$ probability contours in the
  $N=2$ scenario for IH, assuming a putative measurement of SHiP of the two masses $M_1, M_2$ and their mixings to electrons and muons. We assume uncertainties of $1\%, 10\%$ for the masses and mixings in the grey contours and $0.1\%, 1\%$ in the red ones. An additional posterior probability in light blue is shown for a combination of SHiP and a measurement of the phase $\delta$ in DUNE or HyperK. The parameters shown are the observables $Y_B$, $|m_{\beta\beta}|$, $|U_{\tau4}|^2$,  $\delta$ and $\phi_1$.  }
\end{center}
\end{figure}

As a proof of principle we have studied the posterior probabilities for a hypothetical measurement of SHiP of $M_1$ and $M_2$ and their 
respective couplings to electrons and muons for IH. The point chosen is indicated by a star in Fig.~\ref{fig:trianglen2ih}.  We did not look for a very special nor optimal point, simply that it was within the range of SHiP and could explain the baryon asymmetry. The corresponding triangle plots are shown in Fig.~\ref{fig:triangleship} for two values of the assumed errors in this experiment: relative errors on masses and mixings of $(1\%, 10\%)$ and the much more optimistic $(0.1\%, 1\%)$. 
 \begin{figure}[h]
 \begin{center}
 \includegraphics[scale=0.7]{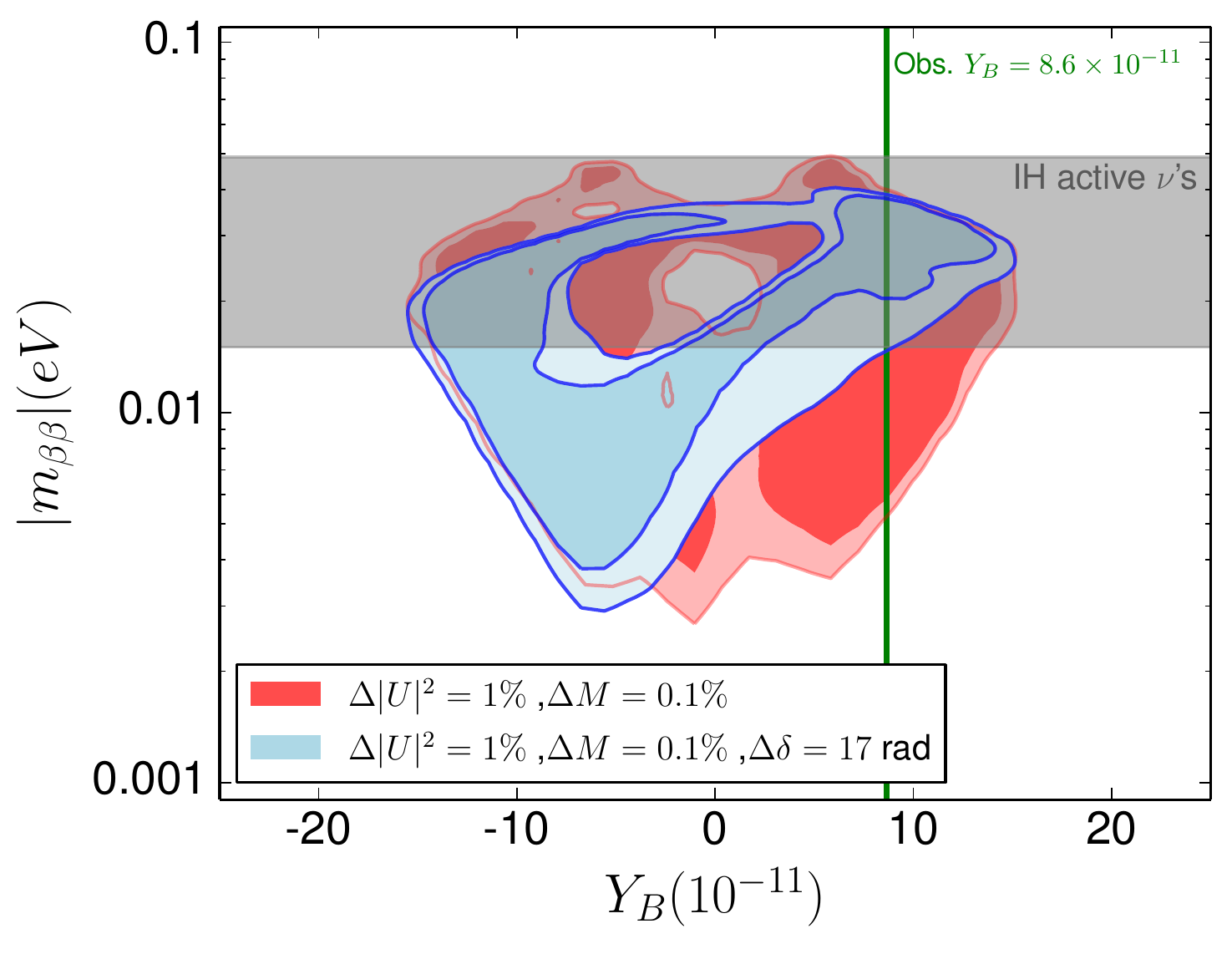} 
\caption{\label{fig:mbbvsYB_SHIP} Posterior probabilities in the $|m_{\beta\beta}|$ vs $Y_B$ plane from a putative measurement at SHiP, assuming
$0.1\%, 1\%$ uncertainties in the masses and mixings (red) or the latter with an additional measurement of $\delta$ in DUNE and HyperK (blue). 
The  grey band is the standard $3\nu$ expectation.}
\end{center}
\end{figure}
We furthermore  considered a combination of SHiP, with the more optimistic errors, with a determination of $\delta$ in future neutrino oscillation experiments  such as 
HyperK and DUNE. We have assumed $\sigma_\delta \simeq 10^\circ$ as derived from the studies in references \cite{Acciarri:2015uup,Abe:2015zbg}. The $|m_{\beta\beta}|$ versus $Y_B$ plot is zoomed in in Fig.~\ref{fig:mbbvsYB_SHIP}.

\begin{figure}[h]
 \begin{center}
 \includegraphics[scale=0.7]{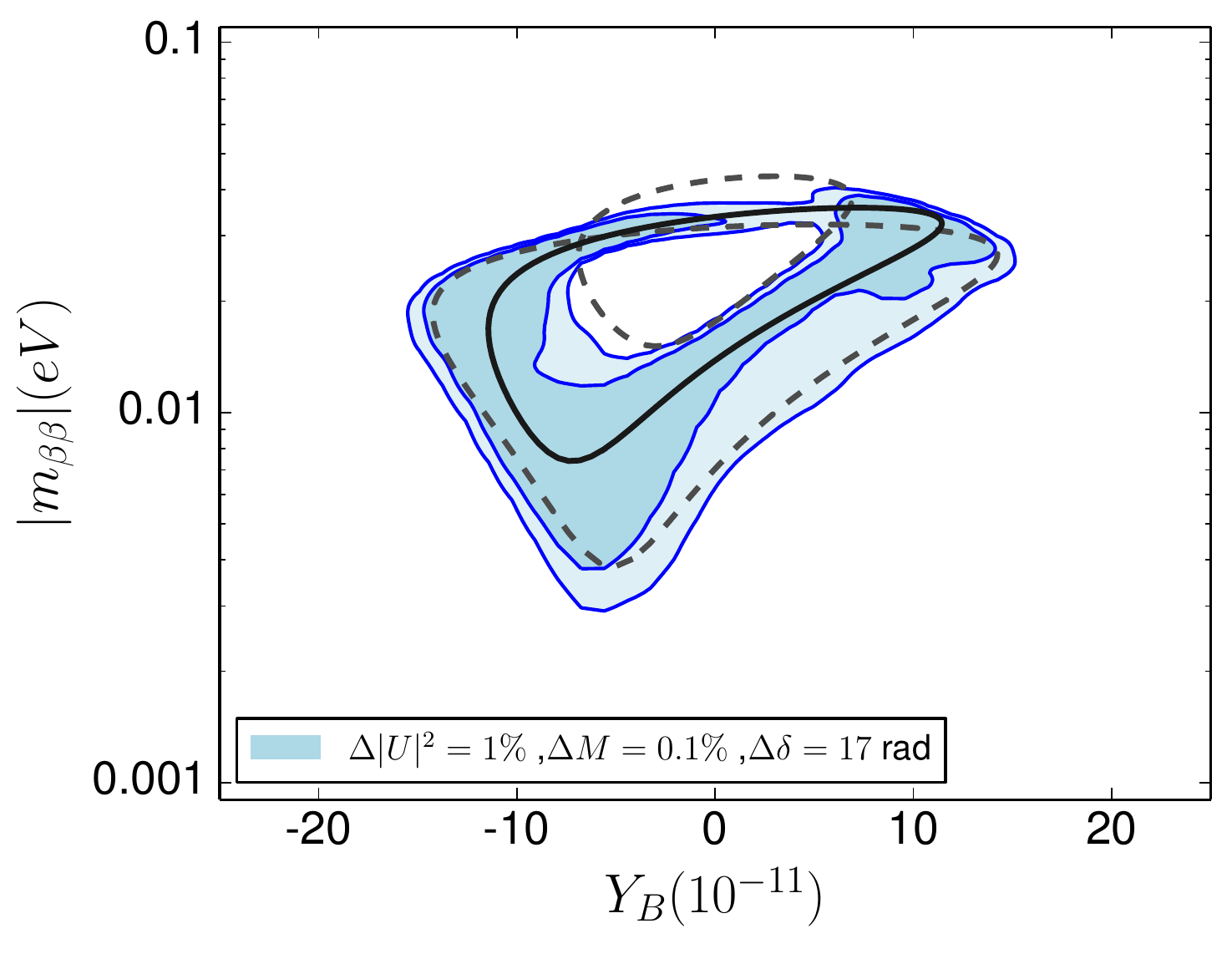} 
\caption{\label{fig:errorplot} Posterior probabilities in the $|m_{\beta\beta}|$ vs $Y_B$ plane from a putative measurement at SHiP, assuming
$0.1\%, 1\%$ uncertainty and an additional measurement of $\delta$ in DUNE and HyperK, together with the contours obtained by changing $\theta$ and every other parameter fixed to the test point values (solid line) or with $\delta =\delta_{\rm test} - \sigma_\delta$ and $\phi_1, \gamma$ in each case tuned to keep $|U_{e4}|^2$, $|U_{\mu4}|^2$ fixed. There are two solutions for $\phi_1$ and these correspond to the two dashed lines. }
\end{center}
\end{figure}

Clearly the determination of $\delta$ results in a more clear correlation between $|m_{\beta\beta}|$ and $Y_B$. In fact the resulting width of the contour can be understood from the propagation of the error in $\delta$ on the determination of $\phi_1$. This is shown in Fig.~\ref{fig:errorplot}, where we compare the posterior
probability with the three curves obtained in the following way: 1) fixing the parameters to those of the test point  and changing only $\theta$ (solid line), 2) the same as 1) except  $\delta$ which is fixed to $\delta_{\rm test} - \sigma_\delta$, and correspondingly $\phi_1, \gamma$ tuned to keep the mixings unchanged.  There are two solutions for $\phi_1$ and the two dashed lines correspond to the two values  (see  Fig.~\ref{fig:phi1vsd}), according to the analytical approximate formulae. The region encompassed by these lines is  roughly similar to the $68\%$ and $90\%$ regions. 

For NH, the expectations are much more pessimistic, since the SHiP measurement would have a hard time to pin down $\phi_1$, even if $\delta$ is known, and therefore two unknowns $\phi_1, \theta$ would remain. They could be determined in principle from a measurement of $Y_B$ and $|m_{\beta\beta}|$ but not from one single measurement and therefore the baryon asymmetry will be very difficult to predict in this case.  The measurement of $|m_{\beta\beta}|$ for NH would nevertheless be futuristic since the value would be roughly a factor 10 smaller, which is beyond the reach of the next generation of neutrinoless double beta decay experiments.

\section{$U_{\rm PMNS}$ phases from SHiP and neutrinoless double beta decay}

We have seen that the ratios of electron and muon mixings that could be measured at SHiP could give very interesting information of the phases of the PMNS matrix. We stress that this is independent of whether the baryon asymmetry can be explained or not. This relies on the fact that the mixings are large enough so that they can be discovered at SHiP. In this case, the ratio of the electron to muon mixings are dominantly  controlled
by the phases of the PMNS mixing matrix, as can be seen in eqs.~(\ref{eq:uihanal}) and (\ref{eq:unhanal}).  

\begin{figure}[h]
 \begin{center}
 \includegraphics[scale=0.6]{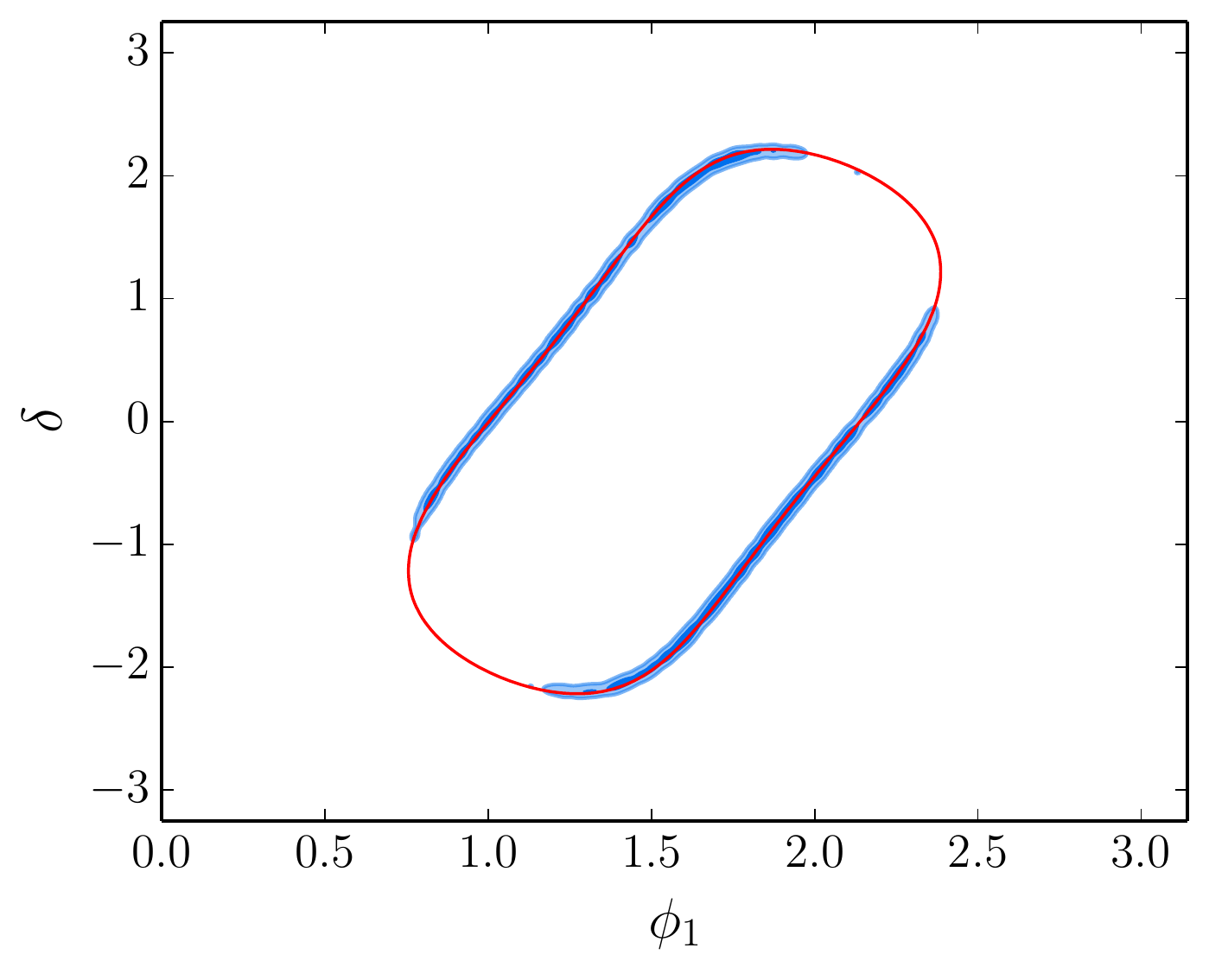} 
\caption{\label{fig:phi1vsd} Posterior probabilities from a SHiP measurement of the masses and mixings with $e, \mu$ on the plane $(\phi_1,\delta)$ compared with the result of the analytical ratio (red line) derived from eqs.~(\ref{eq:uihanal}) for parameters in the test point. }
\end{center}
\end{figure}

In Fig.~\ref{fig:phi1vsd}, we compare the correlation expected on the $(\phi_1,\delta)$ plane from a putative measurement of this ratio using the analytical formulae of eqs.~(\ref{eq:uihanal}) for the test point of the previous section, to  the posterior probabilities obtained with the most competitive SHiP uncertainties assumed in the previous section.  Clearly the analytical formulae work very well and demonstrate the potential of SHiP in constraining the CP violating phases of the mixing matrix. 

This has the following interesting consequence. If neutrinoless double beta decay could be measured with sufficient precision and the effect of the unknown $\theta$ would be small (this would happen for example in a more degenerate situation or for larger heavy masses which suppress the heavy contribution to neutrinoless double beta decay),  the combination of this measurement with that at SHiP could provide information on the phase $\delta$. Quantifying what is the reach of the combination of SHiP and neutrinoless double beta
decay on $\delta$ is very interesting and deserves a dedicated study. 
  
  Reversely, if a measurement of $\delta$ could be obtained from neutrino oscillation measurements, a putative measurement of SHiP could provide a more precise prediction of the neutrinoless double beta decay amplitude if the neutrino ordering is inverted. This would be an extra motivation for improving the nuclear matrix element determination. 
  
\begin{figure}[h]
 \begin{center}
 \includegraphics[scale=0.6]{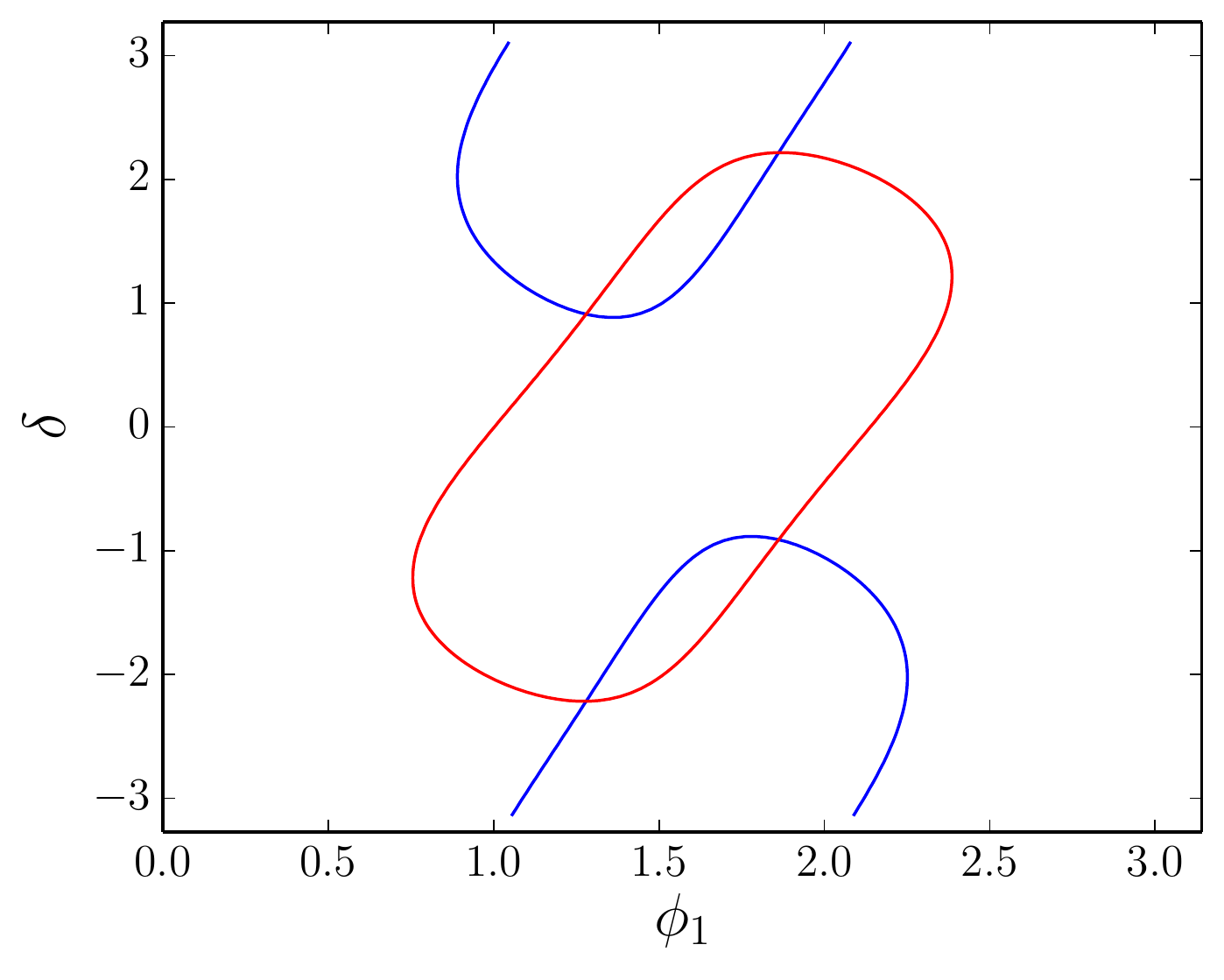} 
\caption{\label{fig:shipmt} Isocurves for the ratios  $|U_{e4}|^2/|U_{\mu 4}|^2$ (red) and $|U_{e4}|^2/|U_{\tau 4}|^2$ (blue) derived from eqs.~(\ref{eq:uihanal}).  }
\end{center}
\end{figure}

  The possibility of measuring also the tau mixing at SHiP could help to resolve the degeneracy. Fig.~\ref{fig:shipmt} shows the isocurves  of $|U_{e4}|^2/|U_{\mu 4}|^2$ and $|U_{e4}|^2/|U_{\tau 4}|^2$ in the $(\phi_1, \delta)$ plane. The test point we used did not have sensitivity to 
  the tau mixing according to the expectations for SHiP, but if this measurement could be improved this would also be useful to reduce the
  $(\delta, \phi_1)$ degeneracy. 
  
\section{Conclusions}

We have studied the production of the matter-antimatter asymmetry in low-scale ${\mathcal O}({\rm GeV})$ seesaw 
models. We have improved  our previous study \cite{Hernandez:2015wna}
by including the washout processes from gauge interactions and Higgs decays and inverse decays, quantum statistics 
in the computation of all rates, as well as spectator processes. This together with a more efficient numerical
 treatment of the equations have allowed us to perform the first bayesian
exploration of the full parameter space that explains the observed baryon asymmetry in the context of the minimal
 model, where only two singlets play a role in the generation of the baryon asymmetry.

We have demonstrated that successful baryogenesis is possible with a mild heavy neutrino degeneracy, and more interestingly that these less fine-tuned solutions prefer smaller masses $M_i \leq 1$GeV, which is the target region 
of SHiP, and significant non-standard contributions to neutrinoless double beta decay.
We have also demonstrated the complementarity of future putative measurements from
SHiP, neutrinoless double beta decay and searches for leptonic CP violation in neutrino oscillations, in the quantitative prediction of the baryon asymmetry within the minimal model. If singlets with masses
in the GeV range would be discovered in SHiP and the neutrino ordering is inverted, the possibility to predict 
the baryon asymmetry (maybe up to a sign) looks in principle viable, in contrast with previous beliefs. Unrelated 
 to whether the baryon asymmetry is explained or not, we have also shown that a measurement of the electron and 
 muon mixings of heavy species within  SHiP range could precisely determine, in the minimal model, a combination 
 of the two phases of the $U_{\rm PMNS}$ matrix.

\begin{acknowledgments}

We warmly thank  N.~Rius for useful discussions and earlier collaboration. 
This work was partially supported by grants FPA2014-57816-P, PROMETEOII/2014/050, and  the European projects 
H2020-MSCA-ITN-2015//674896-ELUSIVES and H2020-MSCA-RISE-2015. PH acknowledges 
the support of the Munich Institute for Astro- and Particle Physics (MIAPP) of the DFG cluster of excellence Origin and Structure of the Universe to participate in the workshop {\it Why is there more matter than antimatter in the Universe}, where this work was first presented prior to publication \cite{talk}.

\end{acknowledgments}

\bibliographystyle{JHEP}
\bibliography{biblio}

\end{document}